\documentclass[twocolumn,english,aps,prb,superscriptaddress,bibnotes,amsmath,amssymb,floatfix]{revtex4-1}
\usepackage[colorlinks=true,citecolor=blue,linkcolor=magenta,pdfstartview=FitH,pdfstartpage=1]{hyperref}
\usepackage[utf8]{inputenc}
\usepackage[english]{babel}
\usepackage[T1]{fontenc}
\usepackage{textcomp}

\usepackage[markup=nocolor, authormarkupposition=left]{changes} 
\usepackage{soul}
\usepackage{url}
\usepackage{makecell}
\setaddedmarkup{{\color{red}#1}}

\usepackage{graphicx}
\usepackage{epstopdf}
\graphicspath{{./figures/}}
\makeatletter

\newcommand{\Rmnum}[1]{\expandafter\@slowromancap\romannumeral #1@}
\makeatother

\begin{document}
\title{Metrology-grade mid-infrared spectroscopy for multi-dimensional perception}

\author{Baoqi Shi}
\thanks{These authors contributed equally to this work.}
\affiliation{International Quantum Academy and Shenzhen Futian SUSTech Institute for Quantum Technology and Engineering, Shenzhen 518048, China}
\affiliation{Department of Optics and Optical Engineering, University of Science and Technology of China, Hefei 230026, China}

\author{Chenxi Zhang}
\thanks{These authors contributed equally to this work.}
\affiliation{International Quantum Academy and Shenzhen Futian SUSTech Institute for Quantum Technology and Engineering, Shenzhen 518048, China}
\affiliation{College of Physics and Optoelectronic Engineering, Shenzhen University, Shenzhen 518060, China}

\author{Ming-Yang Zheng}
\thanks{These authors contributed equally to this work.}
\affiliation{Jinan Institute of Quantum Technology and CAS Center for Excellence in Quantum Information and Quantum Physics, University of Science and Technology of China, Jinan 250101, China}
\affiliation{Hefei National Laboratory, University of Science and Technology of China, Hefei 230088, China}

\author{Yue Hu}
\affiliation{International Quantum Academy and Shenzhen Futian SUSTech Institute for Quantum Technology and Engineering, Shenzhen 518048, China}
\affiliation{Southern University of Science and Technology, Shenzhen 518055, China}

\author{Zeying Zhong}
\affiliation{International Quantum Academy and Shenzhen Futian SUSTech Institute for Quantum Technology and Engineering, Shenzhen 518048, China}
\affiliation{Southern University of Science and Technology, Shenzhen 518055, China}

\author{Zhenyuan Shang}
\affiliation{International Quantum Academy and Shenzhen Futian SUSTech Institute for Quantum Technology and Engineering, Shenzhen 518048, China}
\affiliation{Southern University of Science and Technology, Shenzhen 518055, China}

\author{Wenbo Ma}
\affiliation{Jinan Institute of Quantum Technology and CAS Center for Excellence in Quantum Information and Quantum Physics, University of Science and Technology of China, Jinan 250101, China}

\author{Xiu-Ping Xie}
\affiliation{Jinan Institute of Quantum Technology and CAS Center for Excellence in Quantum Information and Quantum Physics, University of Science and Technology of China, Jinan 250101, China}
\affiliation{Hefei National Laboratory, University of Science and Technology of China, Hefei 230088, China}

\author{Xue Bai}
\affiliation{International Quantum Academy and Shenzhen Futian SUSTech Institute for Quantum Technology and Engineering, Shenzhen 518048, China}
\affiliation{Qaleido Photonics, Shenzhen 518048, China}

\author{Yi-Han Luo}
\affiliation{International Quantum Academy and Shenzhen Futian SUSTech Institute for Quantum Technology and Engineering, Shenzhen 518048, China}

\author{Anting Wang}
\affiliation{Department of Optics and Optical Engineering, University of Science and Technology of China, Hefei 230026, China}

\author{Hairun Guo}
\affiliation{Key Laboratory of Specialty Fiber Optics and Optical Access Networks, Shanghai University, Shanghai 200444, China}

\author{Qiang Zhang}
\affiliation{Jinan Institute of Quantum Technology and CAS Center for Excellence in Quantum Information and Quantum Physics, University of Science and Technology of China, Jinan 250101, China}
\affiliation{Hefei National Laboratory, University of Science and Technology of China, Hefei 230088, China}
\affiliation{Hefei National Research Center for Physical Sciences at the Microscale and School of Physical Sciences, University of Science and Technology of China, Hefei 230026, China}
\affiliation{CAS Center for Excellence in Quantum Information and Quantum Physics, University of Science and Technology of China, Hefei 230026, China}

\author{Junqiu Liu}
\email[]{liujq@iqasz.cn}
\affiliation{International Quantum Academy and Shenzhen Futian SUSTech Institute for Quantum Technology and Engineering, Shenzhen 518048, China}
\affiliation{Hefei National Laboratory, University of Science and Technology of China, Hefei 230088, China}

\maketitle
\noindent\textbf{The mid-infrared spectral window is essential for molecular fingerprinting and atmospheric sensing, yet unlocking its full potential is currently constrained by a fundamental instrumental trade-off: 
existing systems cannot simultaneously deliver broad bandwidth, high photon flux, and metrological frequency fidelity. 
Here, we resolve this bottleneck by demonstrating a metrology-grade spectroscopic system based on difference frequency generation, driven by widely tunable, near-infrared diode lasers traceable to atomic standards. 
Our system achieves continuous tunability across the 3--3.7 \textmu m atmospheric window and delivers output power exceeding 45~mW with an absolute frequency accuracy of 7.2~MHz. 
We harness this convergence to overcome a critical barrier in integrated photonics, unambiguously identifying and eliminating hydrogen-induced absorption in silicon nitride microresonators to achieve an 88-fold reduction in optical loss.
We further reveal multi-phonon absorption in the silica cladding as the fundamental limit to mid-infrared integrated photonics. 
Finally, we demonstrate the system's versatility through scattering-resilient LiDAR capable of penetrating optically dense fog, and dual-modality sensing that simultaneously retrieves target distance and chemical composition. 
By unifying the rigor of frequency metrology with the versatility of broadband sensing, this architecture establishes a new paradigm for multi-dimensional perception in complex environments.
}

The mid-infrared (MIR) spectral region hosts the fundamental vibrational-rotational transitions of molecules, exhibiting absorption cross-sections orders of magnitude larger than their near-infrared (NIR) overtones \cite{Schliesser:12}.
This unique sensitivity underpins transformative applications ranging from planetary-scale greenhouse gas monitoring to non-invasive medical diagnostics \cite{Petersen:14, Diddams:20, Liang:25}. 
In demanding scenarios---such as life-support monitoring in space or sensing in dynamic industrial environments---systems must resolve complex chemical mixtures rapidly and unambiguously \cite{Coddington:16}.  
Consequently, high-performance spectroscopy requires a convergence of capabilities rarely co-existing in a single instrument: 
broad bandwidth for multi-species coverage, fine resolution to distinguish overlapping lines, exact frequency determination,
high spectral brightness, high signal-to-noise ratio (SNR), and real-time processing speed.

\begin{figure*}[t!]
\centering
\includegraphics[width=\linewidth]{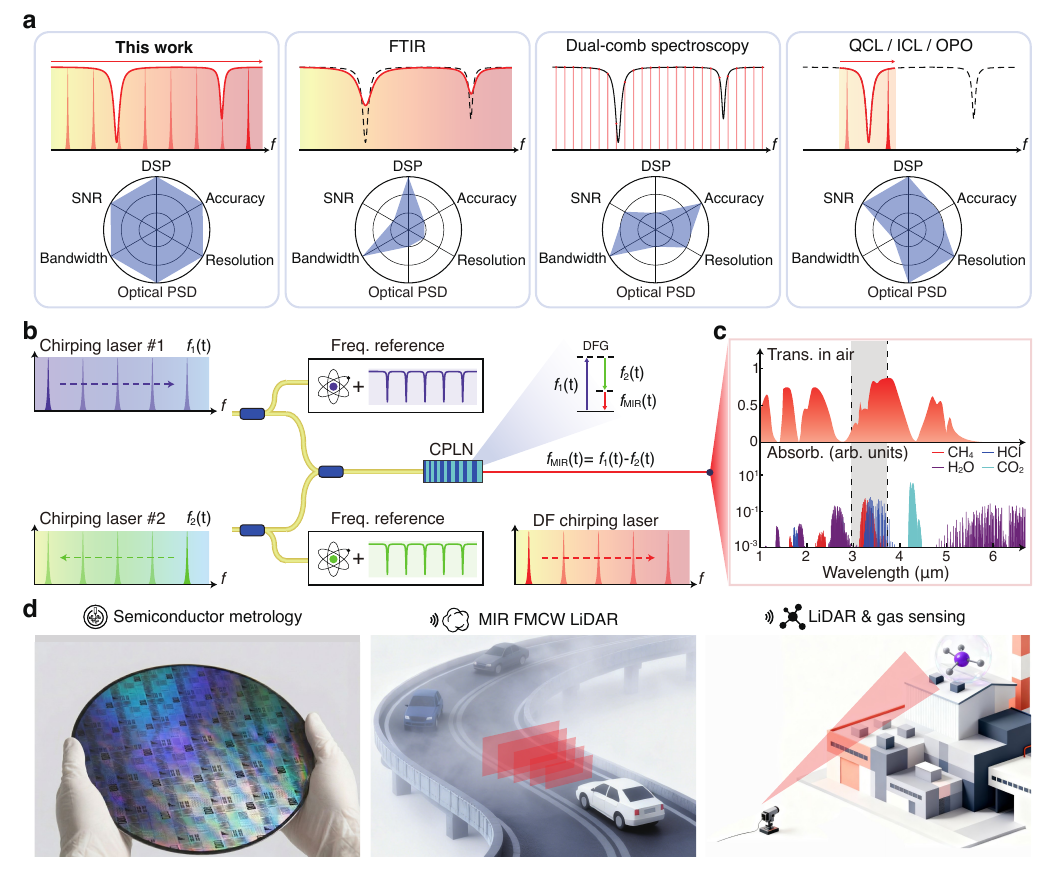}
\caption{
\textbf{Metrology-grade mid-infrared spectroscopic systems breaking the performance trade-offs}.
\textbf{a}. 
Performance benchmarking against established techniques: Fourier transform infrared (FTIR) spectroscopy, dual-comb spectroscopy, and standard tunable CW laser spectroscopy (including QCL, ICL, and OPO). 
The radar chart illustrates how our approach resolves the fundamental instrumental trade-offs, uniquely combining high optical power spectral density (PSD), high SNR, and seamless broad bandwidth with metrological frequency accuracy and precision via simplified digital signal processing (DSP). 
\textbf{b}. 
Schematic architecture. 
A widely tunable, MIR CW laser is synthesized via difference frequency generation (DFG) of two NIR chirping lasers in a chirped periodically poled lithium niobate (CPLN) waveguide. 
The instantaneous frequencies of the NIR pump lasers are rigorously calibrated using fiber cavities and referenced to atomic hyperfine transitions.
\textbf{c}. 
Atmospheric transmission \cite{Musgraves:19} (top) and molecular absorption profiles \cite{Goldenstein:17,Gordon:22} (bottom). 
The system targets the 3--3.7 \textmu m band, exploiting a highly transparent atmospheric window that minimizes H$_2$O and CO$_2$ interference while accessing the strong absorption fingerprints of CH$_4$ and HCl. 
\textbf{d}. 
Emerging applications enabled by our MIR platform: non-destructive semiconductor metrology, scattering-resilient FMCW LiDAR, and simultaneous ranging and gas sensing.
}
\label{Fig:1}
\end{figure*}

Despite this urgent demand, established techniques suffer from fundamental trade-offs, illustrated in Fig.~\ref{Fig:1}a. 
Fourier transform infrared spectroscopy (FTIR) \cite{Hodgkinson:12} offers broad bandwidth but is constrained by low brightness and limited frequency fidelity (i.e., precision and accuracy). 
Conversely, while MIR dual-comb spectroscopy \cite{Ycas:18, Bao:21, Peng:23, Wan:25} enables high-precision, time-resolved measurements, it distributes optical power across tens of thousands of comb lines. 
This results in low optical power spectral density (PSD), severely limiting the photon flux per mode available for long-path remote sensing \cite{Giorgetta:10}.
In contrast, continuous-wave (CW) tunable lasers---such as quantum cascade lasers (QCLs) \cite{Faist:94}, interband cascade lasers (ICLs) \cite{Vurgaftman:15}, or optical parametric oscillators (OPOs) \cite{ZhangZ:20, ZhangZ:20b, Foote:21}---maximize photon flux and interaction lengths that are pivotal for nearly all ultrasensitive optical detection. 
However, these sources are historically hampered by limited mode-hop-free tuning ranges, precluding the seamless, broadband characterization required for complex molecular fingerprinting.

An ideal spectroscopic platform would bridge this gap by combining the coherence and high photon flux of CW lasers with the bandwidth of frequency combs and the absolute accuracy of metrology standards. 
Such a synthesis represents more than an incremental improvement; 
it is a prerequisite for unlocking new frontiers in MIR science outlined in Fig. \ref{Fig:1}d. 
For instance, in semiconductor manufacturing, while MIR light sources are used to monitor chemicals, a spectrometer with absolute frequency accuracy and fine frequency resolution is critical for decoupling fabrication-induced errors from intrinsic material imperfection. 
Similarly, for autonomous driving, standard NIR LiDAR fails in heavy fog due to Mie scattering \cite{Zhang:23b}, while MIR LiDAR capable of penetrating complex environments significantly boosts safety standards.
Ultimately, unifying spectroscopic resolution with spatio-temporal coherence enables a new paradigm: 
fusing spatial geometry and chemical composition into a unified multi-dimensional map. 

Here, we demonstrate a metrology-grade MIR spectroscopic system that simultaneously synergizes broad bandwidth, high brightness, high SNR, and potential for real-time processing. 
As depicted in Fig. \ref{Fig:1}b, our approach leverages difference frequency generation (DFG) driven by two widely tunable, mode-hop-free, NIR external-cavity diode lasers (ECDLs). 
To ensure frequency accuracy, the two pump lasers (spanning 1035--1086 nm and 1536--1580 nm, individually) are rigorously calibrated against fiber cavities and referenced to atomic hyperfine transitions \cite{Preston:96}, yielding a frequency accuracy of 7.2 MHz (see Methods). 
By amplifying and mixing these pumps in a chirped periodically poled lithium niobate (CPLN) waveguide, we generate up to 45.0 mW of MIR radiation.
Details are found in Methods, Extended Data Figs. \ref{ExFig:1} and \ref{ExFig:2}, and Supplementary Materials Notes 1 and 2. 
This configuration delivers an individual mode-hop-free tuning range of 7.49 THz and an aggregate coverage of 19.1 THz (3001--3711 nm). 
Furthermore, we determine the fundamental frequency resolution by characterizing the MIR laser's dynamic linewidth, measuring an average of 242 kHz over 100 \textmu s integration time (see Supplementary Materials Note 3).

We strategically target the 3--3.7 \textmu m atmospheric window \cite{Musgraves:19}---a highly transparent region that minimizes interference from ubiquitous H$_2$O and CO$_2$ while maximizing sensitivity to hydrocarbons (e.g., CH$_4$) and industrial pollutants (e.g., HCl) \cite{Gordon:22}, as illustrated in Fig. \ref{Fig:1}c.
Harnessing this platform, we demonstrate three distinct applications: 
(1) a metrological probe for integrated photonics, which we utilize to isolate and mitigate material-induced optical losses in silicon nitride microresonators; 
(2) a frequency-modulated continuous-wave (FMCW) LiDAR capable of penetrating dense scattering media; 
and (3) a dual-modality sensor that simultaneously retrieves target distance and trace-gas fingerprints from a single waveform.

\begin{figure*}[t!]
\centering
\includegraphics[width=\linewidth]{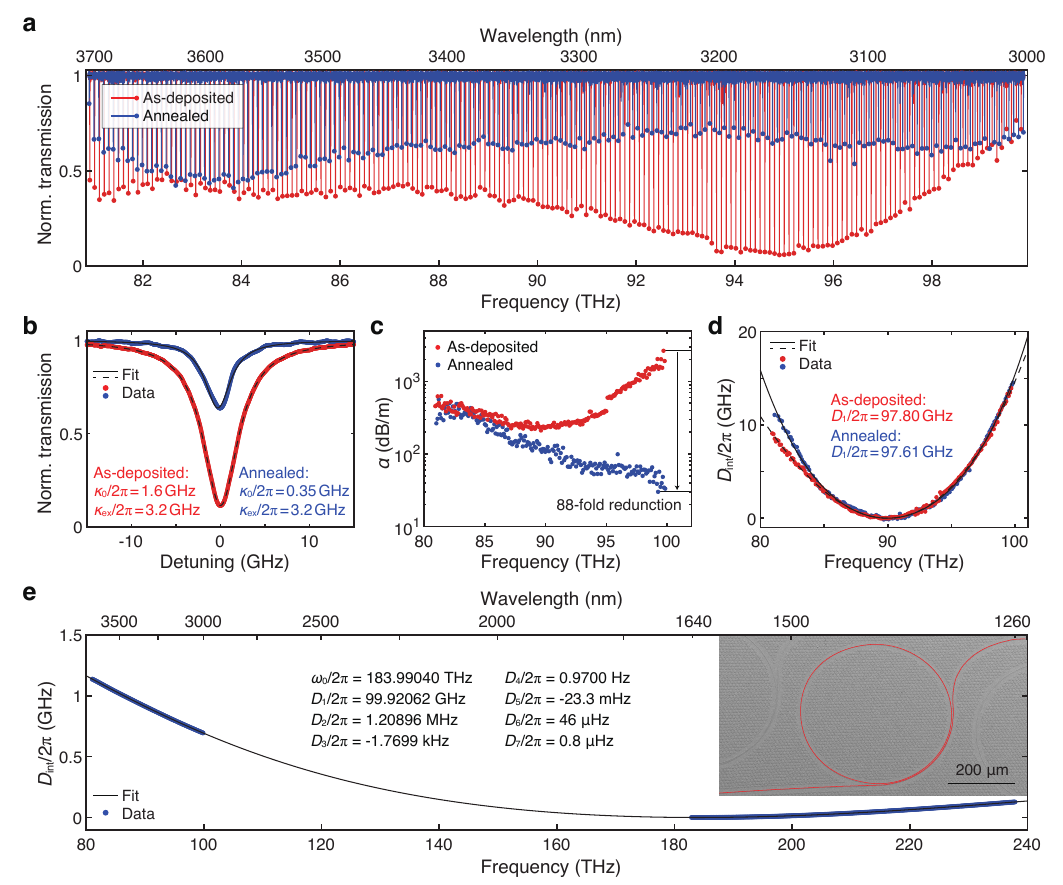}
\caption{
\textbf{Broadband spectroscopic characterization and loss mitigation in silicon nitride microresonators}.
\textbf{a}.
Broadband transmission spectra of the same Si$_3$N$_4$ microresonator before (as-deposited, red) and after (annealed, blue) annealing. 
\textbf{b}.
Comparative line shape analysis of a single resonance at 95.448~THz. 
The high-temperature annealing reduces the intrinsic loss $\kappa_0/2\pi$ from 1.6~GHz (as-deposited, red) to 0.35~GHz (annealed, blue), while the external coupling rate remains constant at $\kappa_\text{ex}/2\pi=3.2$~GHz. 
\textbf{c}.
Frequency-dependent optical propagation loss $\alpha$ converted from $\kappa_0/2\pi$. 
The annealing drastically eliminates the N-H bonds whose absorption peak is located at 100 THz ($\approx3.0$ \textmu m). 
The attenuation $\alpha$ is reduced from 2646~dB~m$^{-1}$ (as-deposited, red) to 30~dB~m$^{-1}$ (annealed, blue)---an 88-fold reduction.  
\textbf{d}.
Integrated dispersion ($D_\text{int}/2\pi$) profiles fitted to a fourth-order polynomial. 
The annealing induces a shift in the free spectral range (FSR, $D_1/2\pi$) from 97.80~GHz to 97.61~GHz at the central frequency $\omega_0/2\pi = 90.011$~THz, reflecting changes in the refractive and group indices. 
\textbf{e}.
Octave-spanning dispersion characterization (1260--3711~nm). 
The profile is obtained by stitching measurements from the NIR (1260--1640~nm, 546 resonances) and MIR (3001--3711~nm, 192 resonances) bands of the same device and fitting to a seventh-order polynomial. 
Inset, false-color SEM image of a Si$_3N_4$ microresonator (waveguide highlighted in red).
}
\label{Fig:2}
\end{figure*}

\vspace{0.3cm}
\noindent \textbf{Mid-infrared integrated photonics}.
Integrated photonics \cite{Thomson:16, Marin-Palomo:17, KimI:21, Shu:22, Yang:22, Guo:23} has matured rapidly in the NIR telecommunication bands, yet the MIR spectral window remains a frontier with immense untapped potential for chemical sensing \cite{Zhou:22} and free-space communications \cite{Dely:21}. 
Recent progress has yielded critical MIR components, including narrow-linewidth Brillouin lasers \cite{Ko:25}, ring lasers \cite{Fuchsberger:25}, microcombs \cite{Wang:13, Yu:16, Meng:21, Opaak:24, Kazakov:25}, and tunable filters \cite{Kazakov:24}. 
Among various material platforms, silicon nitride (Si$_3$N$_4$) is a premier candidate for expanding this ecosystem into the MIR, owing to its ultralow loss and high power handling \cite{Ye:23}. 
However, the scalability of Si$_3$N$_4$ devices beyond 3.5 \textmu m has remained obscured by the onset of multi-phonon absorption in the silicon dioxide (SiO$_2$) cladding \cite{Soref:06b, Kitamura:07, Lin:17}. 
Although this material absorption is well-documented in bulk materials, its specific contribution to waveguide loss---which is governed by the spectral evolution of mode confinement---has not been rigorously quantified.
Consequently, progress has been stalled by the absence of metrology capable of distinguishing manageable process imperfections from fundamental material absorption boundaries.

Unlocking the potential of MIR integrated photonics demands characterization tools that simultaneously offer high frequency accuracy, resolution, and broad spectral coverage. 
Such capabilities are indispensable for quantifying high-quality-factor (high-$Q$) resonances \cite{Vahala:03}, validating dispersion engineering \cite{Okawachi:14}, and aligning resonances with molecular transitions \cite{Griffith:15}. 
Leveraging our system, we report the metrology-grade, broadband characterization of high-$Q$ Si$_3$N$_4$ microresonators, extending their operation into the uncharted 3--3.7 \textmu m band.

Figure \ref{Fig:2}a red trace presents the transmission spectrum of a microresonator based on an as-deposited Si$_3$N$_4$ film from low-pressure chemical vapor deposition (LPCVD). 
As exemplified in Fig. \ref{Fig:2}b, the intrinsic loss $\kappa_0/2\pi$ for each resonance is extracted via Lorentzian fitting (see Methods).  
By converting $\kappa_0/2\pi$ to linear propagation loss $\alpha$ for all probed resonances within the 3--3.7 \textmu m, we uncover a distinct frequency-dependent loss profile as depicted in Fig. \ref{Fig:2}c red dots. 
We note that the localized discontinuity observed near 95 THz arises from parameter indistinguishability at the critical coupling crossover, detailed in Supplementary Note 4.
Crucially, a dramatic increase in $\alpha$ towards 100 THz ($\approx3.0$ \textmu m) is observed, peaking at $\alpha=2646$ dB m$^{-1}$. 
We identify this feature as the absorption fingerprint of residual N-H bonds in the as-deposited Si$_3$N$_4$ film \cite{Lanford:78, Bugaev:12, Luke:15} whose precursors involve ammonia (NH$_3$). 

Guided by this spectral fingerprint, we optimize the fabrication process by adding high-temperature annealing (above 1200$^\circ$C) on this Si$_3$N$_4$ device to eliminate residual hydrogen \cite{Luke:15, Liu:18a}. 
Details are found in Supplementary Materials Note 5. 
Figures \ref{Fig:2}a--c blue traces evidence the efficacy of this mitigation and drastic loss reduction in the annealed device. 
For the mode at $\omega/2\pi=95.448$ THz in Fig. \ref{Fig:2}b, $\kappa_0/2\pi$ decreases from 1.6 GHz to 0.35 GHz, corresponding to $Q_0=\omega/\kappa_0$ improvement to $2.7\times 10^5$.  
Specifically, the value of $\alpha$ at 100 THz drops precipitously from 2646 dB m$^{-1}$ to just 30 dB m$^{-1}$---an 88-fold reduction that confirms the suppression of N-H absorption. 

Eliminating the N-H impurities unmasks a distinct loss profile where attenuation rises as frequency decreases. 
Finite-element simulations rule out geometric leakage (bending and substrate coupling) as the primary driver of this attenuation (see Methods).
At 81 THz ($\approx$3.7 \textmu m), the calculated radiative loss is limited to $\sim$11.4 dB m$^{-1}$---over an order of magnitude lower than the measured value of $\sim$420 dB m$^{-1}$. 
We instead attribute this excess loss to mode delocalization into the absorptive SiO$_2$ cladding \cite{Miller:17}.
Poynting vector analysis of the waveguide geometry reveals that 31\% of the optical power propagates within the cladding at 81 THz, as shown in Extended Data Fig. \ref{ExFig:3}.
Given the onset of strong SiO$_2$ multi-phonon absorption in this spectral window \cite{Soref:06b, Kitamura:07, Miller:17}, this substantial mode overlap is sufficient to account for the observed loss floor.
Consequently, increasing the waveguide width offers a direct pathway to mitigate this limit by enhancing optical confinement.
Characterization results of additional microresonators are found in Supplementary Materials Note~4.

Moreover, our metrology-grade accuracy and precision allow for mapping of microresonator dispersion profile (see Methods). 
Figure \ref{Fig:2}d compares the measured integrated dispersion ($D_\text{int}/2\pi$) profiles for the Si$_3$N$_4$ microresonator before and after the annealing. 
While the waveguide geometry is unchanged, the difference in dispersion is also caused by the N-H absorption via the Kramers-Kronig relations \cite{Lucarini:05, HuY:26}.  

These results establish our system as a powerful, non-destructive diagnostic tool for guiding fabrication process iteration. 
Furthermore, by routing a portion of the NIR pump light, our system enables simultaneous characterization of both NIR and MIR spectral windows. 
Figure \ref{Fig:2}e displays an octave-spanning dispersion profile obtained from one microresonator, stitching data from the NIR (1260--1640 nm) and MIR (3001--3711 nm) bands.
Such broadband capability is critical for validating dispersion engineering \cite{Okawachi:14} in applications requiring octave-spanning spectra \cite{Rao:21}, including supercontinuum generation \cite{Guo:18, Grassani:19, Granger:23}, microcomb generation \cite{Wang:13, Griffith:15, Yu:16, Chen:20} and frequency translation \cite{Foster:06, Li:16}. 
Beyond Si$_3$N$_4$, our spectroscopic system is equally pivotal for advancing other MIR platforms, including chalcogenide glass \cite{Ko:25}, III-V materials \cite{Granger:23, Opaak:24}, Si \cite{Li:11, Yu:16, Miller:17}, Ge \cite{Chang:12, Nedeljkovic:17, SnchezPostigo:21}, and SiGe \cite{Turpaud:24}.

\begin{figure*}[t!]
\centering
\includegraphics[width=\linewidth]{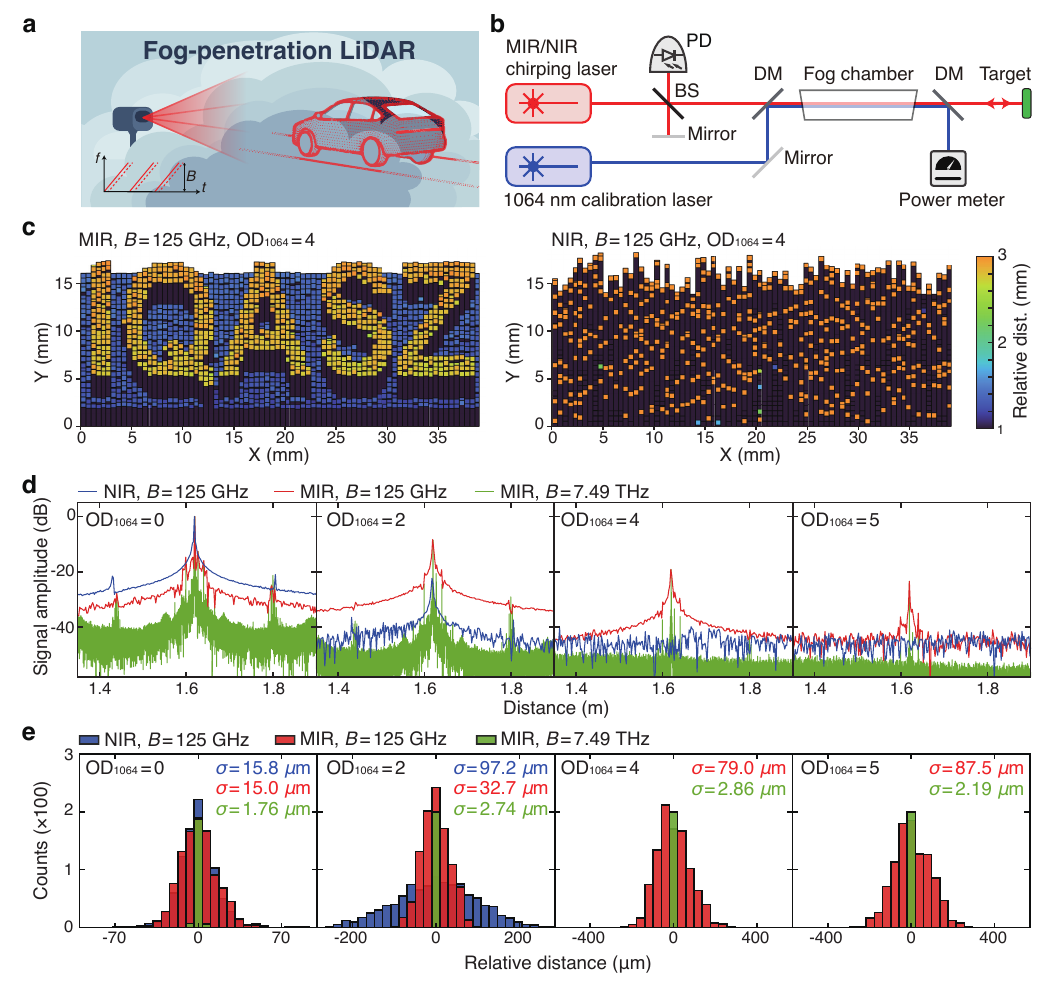}
\caption{
\textbf{Scattering-resilient mid-infrared FMCW LiDAR}.
\textbf{a}.
Illustration of the fog-penetration concept. 
Inset, time-frequency representation of the ranging waveforms with a tuning bandwidth of $B$. 
The time delay between the emitted signal (solid lines) and the received echo (dashed lines) encodes the target distance. 
\textbf{b}.
Experimental testbed. 
A controlled fog chamber simulates scattering environments, with density calibrated in real-time by monitoring the optical density of a co-propagating 1064 nm reference laser (OD$_{1064}$).
PD, photodetector. 
BS, beam splitter. 
DM, dichromatic mirror. 
\textbf{c}.
Depth imaging comparison of a steel target in dense fog (OD$_{1064}=4$) using identical tuning bandwidths ($B=125$ GHz). 
The MIR system successfully resolves the target structure ``IQASZ'', whereas the standard NIR system is completely obscured by noise. 
\textbf{d}.
Signal resilience against scattering. 
The NIR signal (blue, $B=125$ GHz) is extinguished at OD$_{1064} = 4$, while the MIR signal (red, $B=125$ GHz) remains distinguishable even at OD$_{1064}=5$. 
Green traces represent the MIR mode with $B=7.49$~THz, yielding 12 dB suppression of the noise floor compared to the red traces. 
\textbf{e}.
Ranging precision histograms. 
While precision quickly degrades for the narrowband signals (blue and red, $B=125$ GHz) with increasing opacity, the wideband MIR mode (green, $B=7.49$ THz) maintains robust micron-level precision even under OD$_{1064}=5$.  
}
\label{Fig:3}
\end{figure*}

\vspace{0.3cm}
\noindent \textbf{Scattering-resilient ranging}. 
LiDAR is the cornerstone of spatial perception in autonomous systems. 
While FMCW LiDAR offers shot-noise-limited sensitivity and immunity to ambient light \cite{KimI:21, Li:22}, standard NIR (e.g., 1.55 \textmu m) implementations suffer severe attenuation in scattering media like fog or smoke, posing a distinct risk to safety-critical applications \cite{Zhang:23b}.
To address this vulnerability, we exploit the superior propagation characteristics of the MIR spectrum, as shown in Fig. \ref{Fig:3}a. 
Since scattering strength generally scales inversely with wavelength (e.g., via Rayleigh scattering), shifting operation to the 3--3.7 \textmu m atmospheric window significantly reduces attenuation compared to 1.55 \textmu m. 

We validate this penetration capability using a comparative testbed incorporating a controlled fog chamber (see Methods), as shown in Fig. \ref{Fig:3}b.
Scattering environments are simulated via a water mist generator, with the optical density (OD$_{1064}$) quantified in real-time by a co-propagating 1064 nm probe laser. 
To ensure a rigorous comparison, we configure a standard NIR system (centered at 1.55 \textmu m) and our MIR system (centered at 3.56 \textmu m) with identical parameters: 1 mW optical power and $B=125$ GHz tuning bandwidth. 
We compensate for chirping nonlinearity using intrinsic fine frequency calibration. 
Details are presented in Supplementary Materials Note 6.

The resilience of the MIR signal is striking in Fig. \ref{Fig:3}d. 
As scattering density increases, the NIR signal (blue trace) undergoes rapid extinction, falling below the noise floor at OD$_{1064} = 4$. 
In stark contrast, the MIR signal (red trace) retains a robust SNR of 23 dB under identical conditions. 
Even at the dynamic range limit of our detector (OD$_{1064} = 5$), the MIR system maintains a distinguishable SNR of 19 dB. 
Figure \ref{Fig:3}c evidences that this spectral advantage translates directly into imaging performance: 
while the NIR point cloud is completely obscured by noise, the MIR system resolves the depth profile of a steel target at 1.62 m through dense fog with high fidelity. 

Beyond scattering immunity, high-precision ranging demands wide spectral bandwidth. 
Leveraging our system's broad mode-hop-free tuning range, we expand the modulation bandwidth to $B=7.49$ THz. 
This bandwidth not only refines axial resolution ($\delta R \propto c/2B$) but also dramatically improves SNR through Fourier-domain processing gain, which confines signal energy to a narrower spectral bin against broadband noise background. 
The enhancement is highlighted in Fig. \ref{Fig:3}d: 
the $B=7.49$ THz configuration (green trace) yields 12 dB suppression of the noise floor compared to the $B=125$ GHz baseline (red trace).

We calculate the ranging precision $\sigma$ as the standard deviation to the mean, quantified via histogram analysis in Fig. \ref{Fig:3}e. 
Under clear condition OD$_{1064} =0$, both the NIR and narrowband ($B=125$ GHz) MIR systems achieve similar precision of $\sigma\approx15$ \textmu m.
However, as turbidity increases, precision in narrowband systems degrades significantly due to SNR deterioration: 
$\sigma=97.2$ \textmu m for OD$_{1064}=2$ in the NIR, $\sigma=87.5$ \textmu m for OD$_{1064}=5$ in the MIR. 
Conversely, the wideband ($B=7.49$ THz) MIR system demonstrates exceptional resilience, maintaining $\sigma=1.76$ \textmu m precision in clear air and degrading only marginally to $\sigma=2.19$ \textmu m even under extreme attenuation (OD$_{1064}=5$). 
These results establish high-bandwidth MIR FMCW LiDAR as a robust solution for precision ranging in heavily scattering environments. 
By penetrating dense particulate matter, this technology paves the way for all-weather autonomous driving and critical emergency applications such as imaging through smoke-filled buildings or penetrating dense cloud cover to aid search-and-rescue missions. 

\begin{figure*}[t!]
\centering
\includegraphics[width=\linewidth]{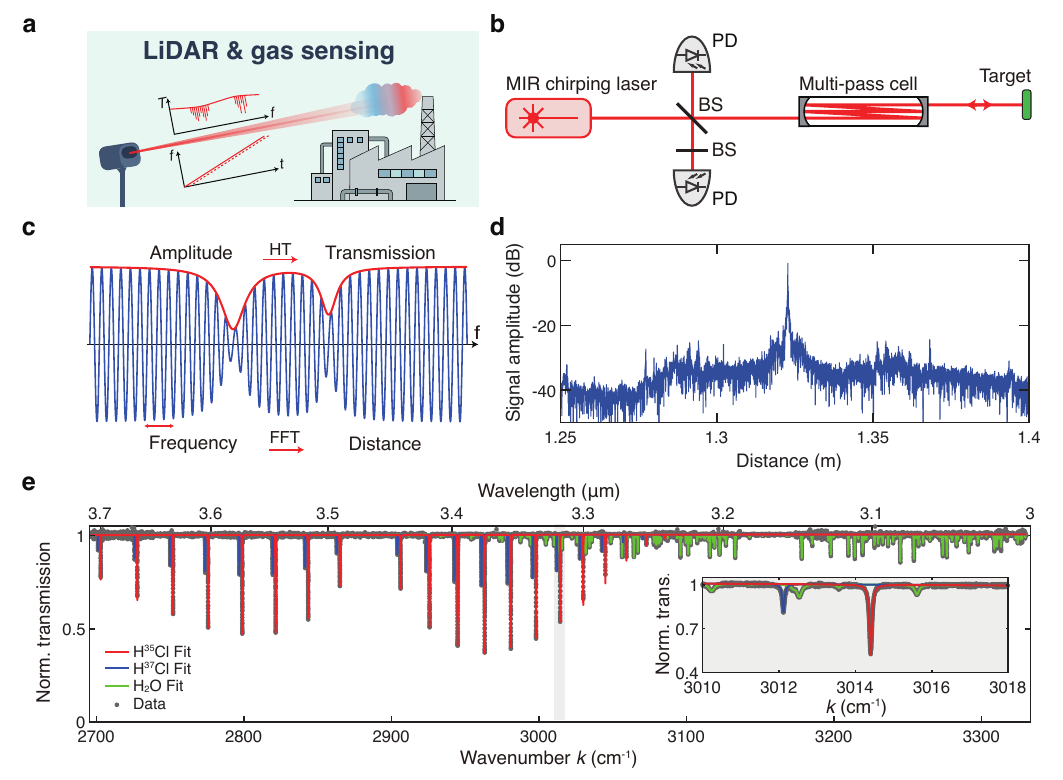}
\caption{
\textbf{Simultaneous ranging and chemical identification from a single optical waveform}. 
\textbf{a}.
Conceptual illustration of the dual-modality perception: 
the system measures the physical distance to a target while simultaneously analyzing the chemical composition of the intervening medium. 
\textbf{b}.
Experimental testbed. 
An industrial emission scenario is simulated using a multi-pass gas cell filled with an HCl mixture.
PD, photodetector. 
BS, beam splitter. 
\textbf{c}.
Signal decoupling principle. 
The interferometric beat signal carries orthogonal information channels: 
range is encoded in the frequency modulation (retrieved via fast Fourier transform, FFT), while spectral fingerprints are encoded in the amplitude envelope (extracted via Hilbert transform, HT).
\textbf{d}.
Reconstructed LiDAR range profile, pinpointing the target location at 1.32 m. 
\textbf{e}.
Quantitative spectroscopic analysis. 
The extracted transmission spectrum resolves the distinct rotational-vibrational transitions of HCl isotopologues (H$^{35}$Cl and H$^{37}$Cl) and background atmospheric H$_2$O vapor. 
Multi-component spectral fitting yields concentrations of 82 ppm for HCl and 0.50\% for H$_2$O.
}
\label{Fig:4}
\end{figure*}

\vspace{0.3cm}
\noindent \textbf{Spectroscopic ranging}. 
The escalating complexity of urban environments necessitates a paradigm shift in sensing technologies---moving beyond simple geometric mapping to multi-dimensional perception that simultaneously captures chemical composition \cite{Kumar:15}. 
To bridge this gap, we exploit our system's 3--3.7 \textmu m coverage to function as both a precision ranger and a metrological gas sensor, as illustrated in Fig. \ref{Fig:4}a. 
This dual capability is vital for monitoring hazardous pollutants such as hydrogen chloride (HCl)---a toxic byproduct of fossil fuel combustion and waste incineration that threatens human health and ecosystem. 

Our approach redefines information extraction from the FMCW LiDAR signal, as shown in Fig. \ref{Fig:4}c.
Conventionally, LiDAR systems extract range information solely from the beat frequency via fast Fourier transform (FFT), typically discarding amplitude variations as noise or fading.
However, in the MIR regime, the signal envelope encodes the spectral absorption ``fingerprint'' of the medium.
To recover this information, we decouple the absorption-induced intensity variations from the frequency-modulated beat signal using Hilbert transform (HT) \cite{Marple:99}.
Algorithmic details are found in Supplementary Materials Note 7.
Crucially, this method leverages the heterodyne gain of coherent detection, where a strong local oscillator amplifies the weak return signal.
This enhancement enables metrology-grade spectroscopic retrieval even from non-cooperative, scattering targets.

We validate this dual-modality perception using a testbed illustrated in Fig. \ref{Fig:4}b. 
The MIR beam is split. 
In the measurement arm it traverses a multi-pass gas cell filled with an HCl mixture before reflecting off a steel target.
The back-reflected beam traverses the cell again before mixing with the beam in the reference arm. 
This setup simulates the remote detection of a pollutant cloud against a solid background. 

Processing a single waveform yields two distinct datasets. 
First, the FFT of the beat signal accurately resolves the target distance at 1.32 m, as shown in Fig. \ref{Fig:4}d.
Second, the amplitude retrieval via HT reconstructs the path-integrated transmission spectrum, as shown in Fig. \ref{Fig:4}e.
The retrieved spectrum agrees with the HITRAN database \cite{Gordon:22}, resolving fine spectral features including distinct HCl isotopologues (H$^{35}$Cl and H$^{37}$Cl) and interference from atmospheric H$_2$O vapor. 
Multi-component fitting \cite{Olivero:77} yields concentrations of 82 ppm for HCl and 0.50\% for H$_2$O, confirming the system's high sensitivity and spectral resolution. 
This capability for continuous, broadband, and metrology-grade measurement is vital for analyzing complex gas mixtures. 
By integrating spectroscopy with LiDAR ranging, our system offers a multi-dimensional perception tool for urban environmental monitoring, enabling not just the detection of hazardous emissions, but their precise localization and identification in three-dimensional space.

\begin{figure*}[t!]
\centering
\includegraphics[width=\linewidth]{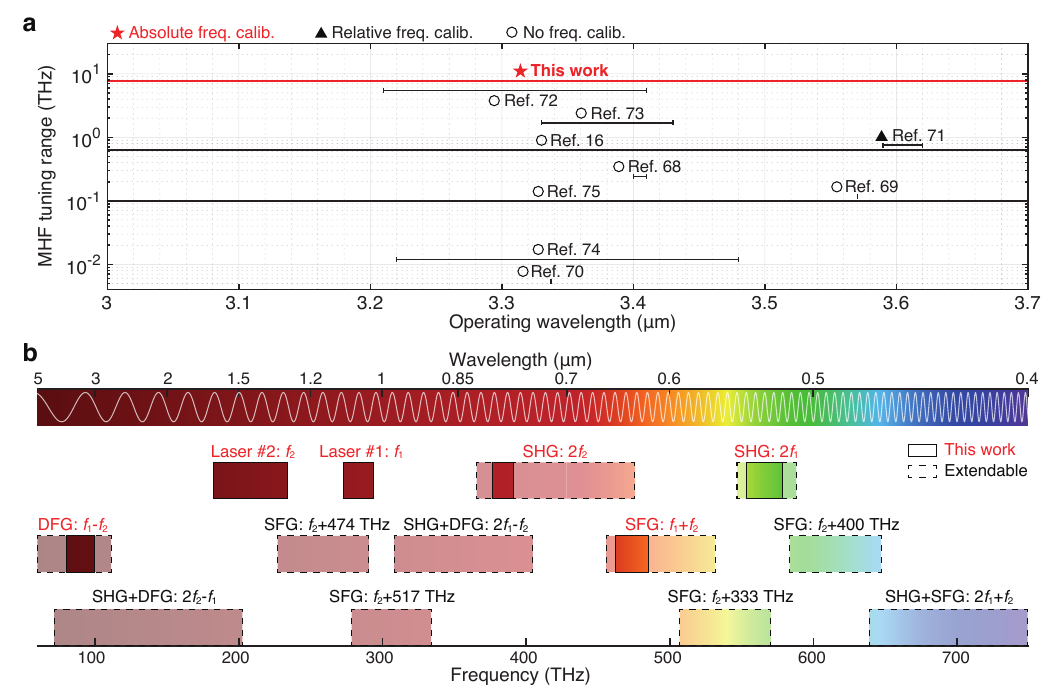}
\caption{
\textbf{Performance benchmarking and scalable multi-spectral synthesis}.
\textbf{a}. 
Comparison of mode-hop-free (MHF) tuning ranges versus operating wavelengths for state-of-the-art CW lasers \cite{Krzempek:13, Forouhar:14, Dong:16, Qu:20, Foote:21, Wang:22b, Gong:23, Dunayevskiy:23, ShahRiyadh:24}. 
Data points are classified by frequency calibration capability: absolute (stars), relative (triangles), and none (circles). 
Our system establishes a new performance benchmark, offering the widest continuous tuning range combined with absolute frequency accuracy. 
\textbf{b}. 
Spectral coverage map spanning the visible to MIR (0.4--5 \textmu m). 
Solid blocks indicate bands experimentally generated in this work, including the primary NIR pumps (1020--1098 nm and 1260--1640 nm) and their nonlinear conversion products: 
SHG (518--541 nm and 766--795 nm), SFG (618--644 nm), and DFG (3001--3711 nm). 
Dashed frames denote theoretically accessible regions via Raman amplification or further mixing with auxiliary sources such as 517 THz optically pumped semiconductor lasers, 474 THz HeNe lasers, 400 THz diode lasers or 333 THz diode lasers. 
This demonstrates the platform's potential for seamless, gap-free spectral synthesis across the entire 0.4--5 \textmu m landscape.
}
\label{Fig:5}
\end{figure*}

\vspace{0.3cm}
\noindent \textbf{Conclusion}. 
We have demonstrated that the historical trade-off between spectral bandwidth, photon flux, and frequency fidelity in the MIR is no longer a fundamental limitation. 
By fusing widely tunable NIR CW lasers with rigorous calibration traceable to atomic standards, our DFG-based architecture provides a deterministic tool for continuous interrogation in the 3--3.7 \textmu m spectral window. 
Benchmarking against state-of-the-art sources listed in Fig. \ref{Fig:5}a, our platform occupies a unique performance space:
it offers the widest continuous tuning coverage combined with absolute frequency accuracy, essentially democratizing metrology-grade spectroscopy.
The transformative potential of this platform is evidenced by its immediate impact across distinct fields. 
In integrated photonics, the system overcomes the critical spectral limits of Si$_3$N$_4$ by pinpointing and eliminating hydrogen-induced absorption. 
This elimination of extrinsic impurities unmasks multi-phonon absorption in the silica cladding as the fundamental barrier, establishing enhanced mode confinement as the critical pathway to unlock the platform's full MIR potential.
Simultaneously, in free-space sensing, it bridges the gap between spatial ranging and chemical analysis, enabling LiDAR systems that penetrate opaque scattering media while simultaneously identifying hazardous pollutants within a single optical waveform.

Looking forward, this architecture is inherently scalable. 
As illustrated in Fig. \ref{Fig:5}b, we have achieved measurement bandwidths spanning the primary NIR pumps ($1020$--$1098$ nm and $1260$--$1640$ nm) and their nonlinear conversion products---including second-harmonic generation (SHG: $518$--$541$ nm and $766$--$795$ nm, see Supplementary Materials Note 1), sum-frequency generation (SFG: $618$--$644$ nm, see Supplementary Materials Note 8), and DFG ($3001$--$3711$ nm)---covering visible to MIR. 
The current system represents just one node in a broader spectral synthesis scheme. 
By employing different combinations of nonlinear processes, the same NIR pump lasers can be converted to cover the 0.4--5 \textmu m spectrum.
This paves the way for a unified, ultra-broadband spectroscopic engine capable of accessing electronic, vibrational, and rotational transitions across the entire molecular landscape. 
Ultimately, by converging metrological frequency fidelity with the versatility of nonlinear optics, our work establishes a new foundation for multi-dimensional perception, unlocking the full information capacity of the MIR spectrum.

\vspace{0.3cm}
\noindent \textbf{Methods}

\setcounter{figure}{0}
\renewcommand{\figurename}{Extended Data Figure}
\begin{figure*}[t!]
\centering
\includegraphics[width=\linewidth]{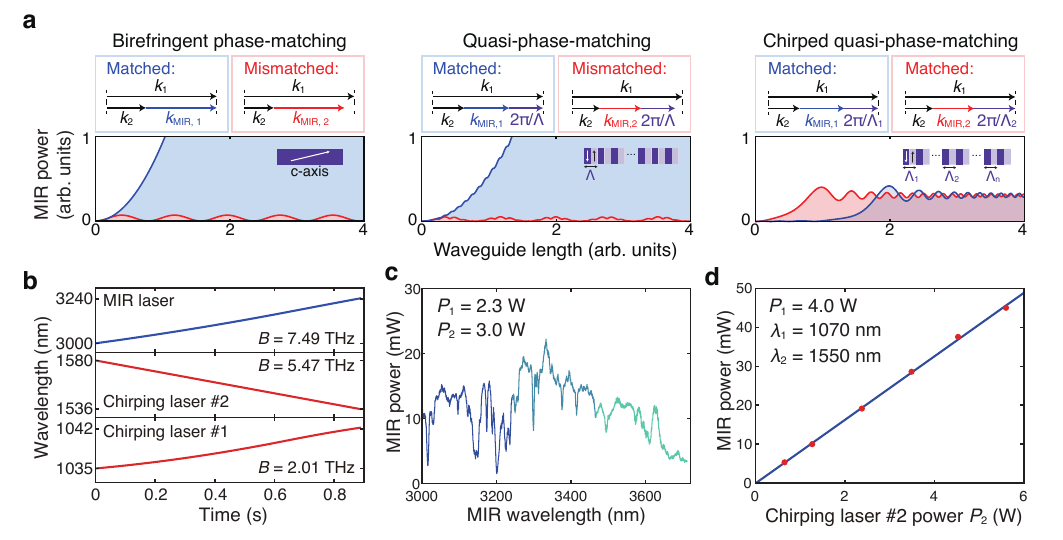}
\caption{
\textbf{Phase-matching mechanisms and broadband mid-infrared synthesis performance}.
\textbf{a}.
Comparison of birefringent phase-matching (BPM), quasi-phase-matching (QPM), and chirped quasi-phase-matching (CQPM). 
Top: wave-vector ($k$) diagrams; 
Bottom: MIR power evolution along the waveguide. 
Unlike BPM and uniform QPM, which exhibit monotonic power growth only at a single phase-matched wavelength (blue regions) and oscillatory behavior elsewhere (red regions), CQPM employs a spatially varying poling period $\Lambda(z)$ to localize phase-matching conditions at different longitudinal positions $z$, enabling broadband DFG.
\textbf{b}.
Real-time frequency tuning profiles calibrated via fiber cavities and atomic hyperfine transitions. 
Chirping laser \#1 scans from 1035 nm to 1042 nm (2.01 THz bandwidth), and chirping laser \#2 scans from 1580 nm to 1536 nm (5.47 THz bandwidth).
The counter-directional chirping of the NIR lasers synthesizes a continuous, mode-hop-free MIR tuning bandwidth of 7.49 THz (3001--3243 nm) in 0.89 s. 
\textbf{c}.
Full spectral coverage. 
Step-tuning the start wavelength of laser \#1 (1035, 1057, and 1079 nm) enables seamless coverage from 3001 nm to 3711 nm. 
These tuning curves are characterized with laser \#1 power $P_1=2.3$ W and laser \#2 power $P_2=3.0$ W.
\textbf{d}.
Power transfer characteristics. 
MIR output power versus pump power $P_2$ with fixed $\lambda_1=1070$ nm, $\lambda_2=1550$ nm, and $P_1=4.0$ W. 
A maximum MIR power of 45.0 mW is generated at $P_2=5.6$ W.
}
\label{ExFig:1}
\end{figure*}

\noindent \textbf{Broadband mid-infrared synthesis}. 
The MIR light is generated via DFG in a CPLN waveguide. 
To overcome the bandwidth constraints of traditional phase-matching, we employ a chirped poling strategy. 
In birefringent phase-matching (BPM) and uniform quasi-phase-matching (QPM) \cite{Yamada:93, Wang:18b}, momentum conservation $\Delta k=0$ is strictly satisfied only for a narrow wavelength range. 
As illustrated in the wave-vector ($k$) diagrams in Extended Data Fig. \ref{ExFig:1}a, a fixed grating vector $2\pi/\Lambda$ cannot compensate for the phase mismatch induced by pump detuning, resulting in oscillatory power transfer rather than cumulative growth. 
In contrast, our device utilizes a chirped quasi-phase-matching (CQPM) profile where the poling period $\Lambda(z)$ varies continuously along the propagation axis $z$. 
The spatially dependent grating vector $2\pi/\Lambda(z)$ ensures that for any pump pair ($k_1$ and $k_2$) within the design bandwidth, the phase-matching condition is spatially localized to a specific longitudinal position. 
Consequently, distinct spectral components initiate conversion at different positions along the waveguide while maintaining high output power. 

To synthesize the broadband MIR laser, chirping laser \#1 scans from 1035 nm to 1042 nm (2.01 THz), while chirping laser \#2 scans in the opposing direction from 1580 nm to 1536 nm (5.47 THz). 
This configuration synthesizes an MIR scan from 3001 nm to 3243 nm, achieving a mode-hop-free tuning bandwidth of 7.49 THz in 0.89 seconds, as depicted in Extended Data Fig. \ref{ExFig:1}b. 
By discretely stepping the start wavelength of laser \#1 (1035, 1057, and 1079 nm), the MIR output achieves seamless coverage from 3001 nm to 3711 nm, as shown in Extended Data Fig. \ref{ExFig:1}c.
The total time for full bandwidth scanning is under 5 seconds. 
Extended Data Fig. \ref{ExFig:1}d shows that the system delivers a maximum output power of 45.0 mW with pump parameters $\lambda_1=1070$ nm ($P_1=4.0$ W) and $\lambda_2=1550$ nm ($P_2=5.6$ W).

\begin{figure*}[t!]
\centering
\includegraphics[width=\linewidth]{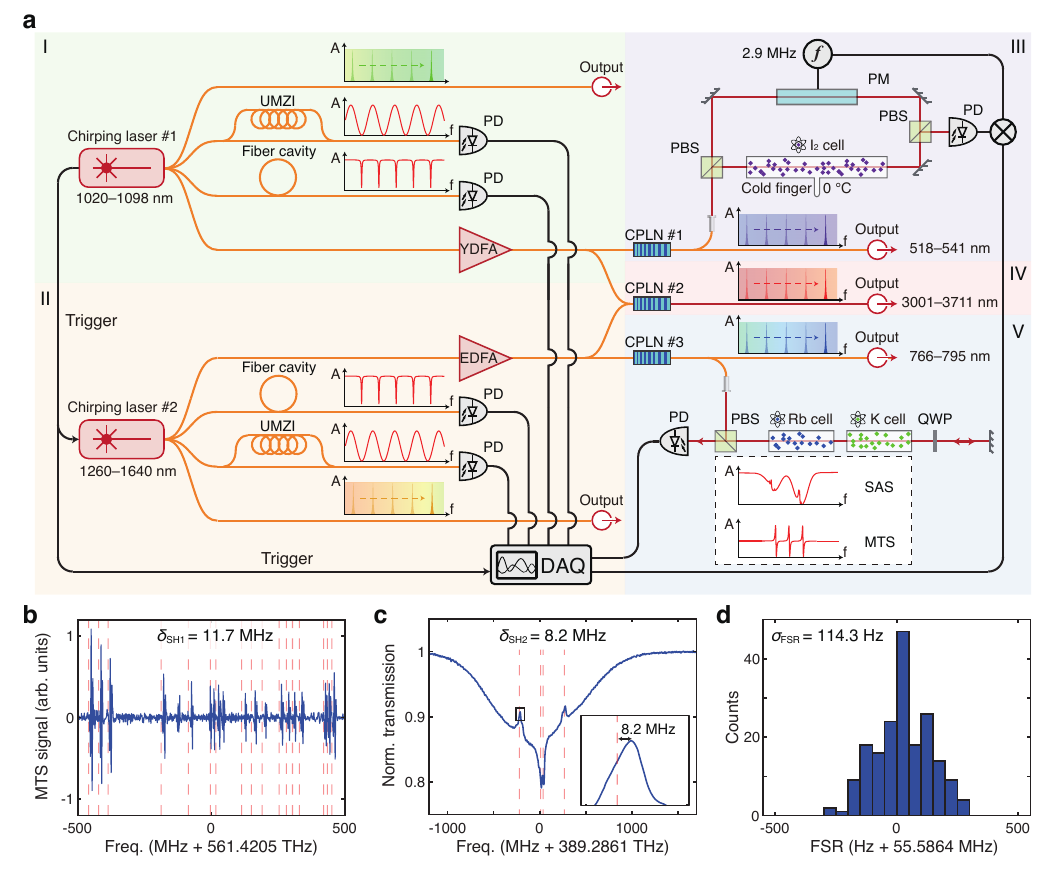}
\caption{
\textbf{Experimental setup and frequency metrology performance}.
\textbf{a}.
Schematic of the spectroscopic system, functionally divided into five regions: 
Regions \Rmnum{1} and \Rmnum{2} track relative frequency calibration of the two NIR pump lasers via fiber cavities and UMZIs; 
Regions \Rmnum{3} and \Rmnum{5} establish absolute frequency referencing via frequency doubling to the visible/near-visible bands; 
and Region \Rmnum{4} generates the MIR DFG output. 
PD, photodetector; 
PBS, polarizing beam splitter;
QWP, quarter-wave plate;
DAQ, data acquisition;
PM, phase modulator.
Bottom-right inset illustrates the principle of saturated absorption spectroscopy (SAS) and modulation transfer spectroscopy (MTS).
\textbf{b}.
Absolute frequency validation for chirping laser \#1 (Region \Rmnum{3}, 518--541 nm). 
Resolved hyperfine transitions of I$_2$ R(53)31-0 line, measured via MTS, shows a residual RMS deviation of  $\delta_\text{SH1}=11.7$ MHz relative to reference data (red dashed lines). 
\textbf{c}.
Absolute frequency validation for chirping laser \#2 (Region \Rmnum{5}, 766--795 nm). 
Resolved hyperfine transitions of K D1 line, measured via SAS, show a residual RMS deviation of $\delta_\text{SH2}=8.2$ MHz relative to reference data (red dashed lines). 
\textbf{d}.
Long-term stability of the fiber cavity's FSR. 
Histogram of FSR values measured at 1490 nm over a 24 hours indicates a standard deviation of $\sigma_{\text{FSR}} = 114.3$ Hz.
}
\label{ExFig:2}
\end{figure*}

\vspace{0.3cm}
\noindent \textbf{Spectroscopic system setup and relative frequency tracking}.
The experimental architecture, detailed in Extended Data Fig. \ref{ExFig:2}a and Supplementary Note 1, is designed for simultaneous multi-spectral operation across five distinct regions (\Rmnum{1}--\Rmnum{5}).
Regions \Rmnum{1} (1020--1098 nm) and \Rmnum{2} (1260--1640 nm) represent the fundamental outputs of the two NIR ECDLs (chirping lasers \#1 and \#2).
These NIR seed sources drive the generation of three derivative bands via nonlinear frequency conversion: 
visible output (Region \Rmnum{3}, 518--541 nm, via SHG), MIR output (Region \Rmnum{4}, 3001--3711 nm, via DFG), and near-visible output (Region \Rmnum{5}, 766--795 nm, via SHG).
Although the NIR seed lasers possess broad tuning ranges, the system's effective operational bandwidth is practically confined by the gain profiles of the YDFA (1035--1086 nm) and the EDFA (1532--1590 nm).
Consequently, only spectral components residing within these gain windows achieve the power levels necessary to pump the CPLN waveguides, where CQPM conditions further select the efficient conversion bands.
This configuration specifically maps the amplified fundamental NIR scans to the target frequencies as follows: 
1036--1082 nm to 518--541 nm (Region \Rmnum{3}); 
1532--1590 nm to 766--795 nm (Region \Rmnum{5}); 
and combined pumps of 1035--1086 nm and 1536--1580 nm to 3001--3711 nm (Region \Rmnum{4}).

In Regions \Rmnum{1} and  \Rmnum{2}, we perform relative frequency calibration by referencing the chirping lasers \#1 and \#2 to fiber cavities. 
The method is illustrated in Ref. \cite{Luo:24} and details are found in Supplementary Materials Note 1. 
The fiber cavities' FSR values (32.32 MHz for laser \#1 and 55.58 MHz for laser \#2) are precisely characterized as relative frequency rulers.
By recording and counting the number of fiber resonances passed by the chirping laser, the relative frequency excursion of the laser is calculated. 
While the fiber resonances provide discrete frequency markers, sub-FSR instantaneous frequencies are resolved by interpolating interference fringes from an unbalanced Mach-Zehnder interferometer (UMZI) \cite{Ahn:05}.
Using the Hilbert transform \cite{Marple:99} to extract the instantaneous phase $\phi(t)$, we calculate the phase difference $\Delta \phi$ between adjacent sampling points. 
The corresponding frequency excursion $\Delta f$ is then derived as:
\begin{equation}
  \Delta f = \frac{\text{FSR}}{\phi_\text{FSR}} \cdot \Delta \phi,
\label{Eq.UMZI}
\end{equation}
where $\phi_\text{FSR}$ is the phase change corresponding to one cavity FSR.

\vspace{0.3cm}
\noindent \textbf{Atomic referencing for absolute frequency traceability}.
We frequency-double the fundamental NIR lasers via CPLN waveguides to access atomic reference lines.
For chirping laser \#1, SHG targets the 518--541 nm band (Region \Rmnum{3} in Extended Data Fig. \ref{ExFig:2}a), where iodine molecules (I$_2$) provide a dense manifold of hyperfine transitions.
We employ modulation transfer spectroscopy \cite{Shirley:82, Cheng:19} to resolve these transitions with sub-megahertz precision, utilizing a counter-propagating pump-probe configuration to generate error signals with high discriminator slopes (see Supplementary Materials Note 1 for details).
To establish absolute frequency references, we utilize the hyperfine transitions of the I$_2$ R(56)32-0, R(53)31-0, and P(28)30-0 lines, for which experimental frequency values are available~\cite{Arie:93, Cheng:19, Shie:13} and are provided in Supplementary Materials Note 1.
The absolute frequency axis is anchored to the I$_2$ R(56)32-0 line ($f_{\text{ref,1}} = 563.2600$ THz)~\cite{Arie:93}. 
We use the hyperfine transitions of the I$_2$ P(28)30-0 line ($f_{\text{ref,2}} = 560.1518$~THz) to validate the relative frequency calibration derived from the fiber cavity. 
We observe a root-mean-square (RMS) deviation of $\Delta_1=-35.8$ MHz between our measured values and those in ref.~\cite{Shie:13}, which we attribute to residual dispersion in the fiber cavity's FSR. 
Consequently, we apply a correction to the second-harmonic (SH) laser frequency using:
\begin{equation}
f'_{\text{SH1}} = \frac{f_{\text{SH1}} - f_{\text{ref,1}}}{1 + \epsilon_1} + f_{\text{ref,1}}
\label{Eq.Correct}
\end{equation}
where $\epsilon_1 = \Delta_1/(f_{\text{ref,1}} - f_{\text{ref,2}})$.
Applying Eq. \ref{Eq.Correct} to the measurement of the I$_2$ R(53)31-0 transitions ($f_{\text{ref,3}} = 561.4205$ THz), as shown in Extended Data Fig. \ref{ExFig:2}b, yields a residual RMS deviation of 11.7 MHz relative to literature values \cite{Cheng:19}. 
This determines the SH laser accuracy as $\delta_\text{SH1}\approx 11.7$ MHz, corresponding to a fundamental frequency accuracy for chirped laser \#1 of $\delta_{\text{laser1}} = \delta_{\text{SH1}}/2 \approx 5.9 $ MHz.

For chirping laser \#2, SHG targets the 766--795 nm band (Region \Rmnum{5} in Extended Data Fig. \ref{ExFig:2}a), where Rubidium (Rb) and Potassium (K) atoms provide distinct hyperfine transitions.
We employ saturated absorption spectroscopy to resolve these transitions with sub-megahertz precision (see Supplementary Materials Note 1 for details).
To establish absolute frequency references, we utilize the hyperfine transitions associated with the Rb D2, K D1, and K D2 lines, whose frequencies have been experimentally measured~\cite{Steck:23,Tiecke:19} and are provided in Supplementary Materials Note 1.
The absolute frequency axis is anchored to the Rb D2 line ($f_{\text{ref,4}} = 384.2286$ THz)~\cite{Steck:23}.
The hyperfine transitions associated with the K D2 line ($f_{\text{ref,5}} = 391.0165$ THz)~\cite{Tiecke:19} are used to examine the accuracy of our relative frequency calibration with the fiber cavity.
An RMS deviation of $\Delta_2=28.4$ MHz is found between our measured values and the values in ref. \cite{Tiecke:19}.
We further correct the SH laser's frequency as:
\begin{equation}
f'_{\text{SH2}} = \frac{f_{\text{SH2}} - f_{\text{ref,4}}}{1 + \epsilon_2} + f_{\text{ref,4}}
\label{Eq.Correct2}
\end{equation}
where $\epsilon_2 = \Delta_2/(f_{\text{ref,4}} - f_{\text{ref,5}})$.
Applying Eq. \ref{Eq.Correct2} to the measurement of the K D1 line ($f_{\text{ref,6}} = 389.2861$ THz), as shown in Extended Data Fig. \ref{ExFig:2}c, yields a residual RMS deviation of 8.2 MHz relative to literature values \cite{Tiecke:19}.
This determines the SH laser accuracy as $\delta_\text{SH2}\approx 8.2$ MHz, corresponding to a fundamental accuracy for chirping laser \#2 of $\delta_{\text{laser2}} = \delta_{\text{SH2}}/2 \approx 4.1$ MHz.

\vspace{0.3cm}
\noindent \textbf{Quantification of accuracy and precision}.
In Extended Data Fig. \ref{ExFig:2}a Region \Rmnum{4}, the MIR laser is generated via DFG.
The absolute frequency accuracy of the MIR laser is determined by the uncorrelated uncertainties of the pump lasers:
\begin{equation}
    \delta_{\text{MIR}} = \sqrt{\delta_{\text{laser1}}^2 + \delta_{\text{laser2}}^2} \approx 7.2 \text{ MHz}.
\label{Eq.Accu}
\end{equation}
The system's precision is ultimately governed by the stability of the fiber cavity FSR.
As presented in Extended Data Fig. \ref{ExFig:2}d, long-term monitoring at 1490 nm over 24 hours reveals an FSR standard deviation of $\sigma_{\text{FSR}} = 114.3$ Hz.
Over the full 5.47 THz tuning range of chirping laser \#2 ($N \approx 10^5$ FSR steps), the cumulative frequency uncertainty scales as $\sigma_{\text{cumulative}} = \sigma_{\text{FSR}} \sqrt{N} \approx 36.1 \text{ kHz}$.
For chirping laser \#1, the narrower tuning range (2.01 THz) implies a correspondingly smaller cumulative error contribution.
Since the MIR frequency is synthesized via DFG, the aggregate frequency uncertainty remains dominated by laser \#2 and is significantly smaller than the dynamic linewidth of the MIR laser (242 kHz, see Supplementary Materials Note 3).
Consequently, the system's precision is limited by the laser linewidth rather than the frequency calibration.

\vspace{0.3cm}
\noindent \textbf{Microresonator dispersion and loss characterization}.
We extract integrated dispersion $D_\text{int}$ by measuring and fitting resonance frequencies to the polynomial:
\begin{equation}
    D_\text{int}(\mu)=\omega_\mu-\omega_0-D_1\mu=\sum_{n=2}^{\cdots}\frac{D_n\mu^n}{n!}
\label{Eq.Dint}
\end{equation}
where $\omega_{\mu}/2\pi$ is the frequency of the $\mu$-th mode relative to the reference mode $\omega_0/2\pi$, 
$D_1/2\pi$ is the microresonator FSR, 
$D_2$ describes group velocity dispersion (GVD), 
and $D_3, D_4$ etc describe higher-order dispersion terms. 
The group index is derived as $n_g = c/(D_1L)$, with $L=1.429$ mm being the physical cavity length of the microresonator in Fig. \ref{Fig:2}. 
For loss quantification, resonances are fitted to a Lorentzian model \cite{Aspelmeyer:14, Pfeiffer:18}:
\begin{equation}
    T(\omega) = 1 - \frac{4\kappa_\text{ex}\kappa_0}{\Delta\omega^2 + (\kappa_\text{ex}+\kappa_0)^2}
\label{Eq.Res}
\end{equation}
where $\Delta\omega/2\pi$ is the laser detuning, 
$\kappa_0/2\pi$ is the intrinsic loss rate, 
and $\kappa_\text{ex}/2\pi$ is the external coupling strength. 
The optical propagation loss (in dB m$^{-1}$) in the waveguide is calculated as $\alpha \approx (27.27 \cdot n_g) / (\lambda \cdot Q_0)$, where $Q_0 = \omega/\kappa_0$ is the intrinsic $Q$-factor.

\begin{figure}[t!]
\centering
\includegraphics[width=\linewidth]{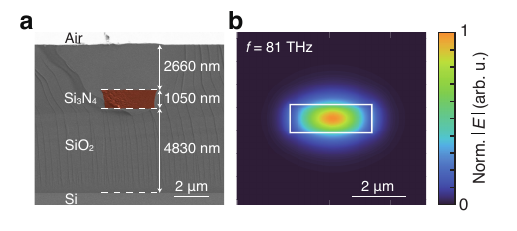}
\caption{
\textbf{Waveguide geometry and modal confinement}.
\textbf{a}.
Cross-sectional false-color SEM image of the Si$_3$N$_4$ waveguide encapsulated by SiO$_2$ cladding. 
The Si$_3$N$_4$ core is highlighted in red.
\textbf{b}.
Simulated fundamental TE$_{00}$ mode profile at 81 THz ($\lambda \approx 3.7$ \textmu m), displaying the normalized electric field magnitude $|E|$. 
}
\label{ExFig:3}
\end{figure}

\vspace{0.3cm}
\noindent \textbf{Finite-element simulations of geometric leakage and optical confinement}.
To quantify radiative losses arising from waveguide bending and substrate leakage, we perform finite-element method (FEM) simulations in a cylindrical coordinate system.
Extended Data Fig. \ref{ExFig:3}a shows the measured SEM of the waveguide cross-section, which is subsequently used to model the FEM simulation. 
The computational domain models the Si$_3N_4$ waveguide (width $w = 3000$ nm, height $h = 1050$ nm) with SiO$_2$ cladding on a silicon (Si) substrate, with boundaries extended 1 \textmu m into the air and Si regions to ensure sufficient field decay.
To isolate geometric leakage from material attenuation, we employ lossless material models utilizing only the real parts of the refractive indices for Si$_3$N$_4$, SiO$_2$, and Si.

We evaluate the radiative leakage via the imaginary part of the effective refractive index, $n''$.
At 81 THz, simulation for a bending radius of $r = 40$ \textmu m yields $n'' \approx 3.5 \times 10^{-6}$.
Increasing the radius to 150 \textmu m significantly reduces this value to $n''\approx 7.8 \times 10^{-7}$.
Further expansion to $r = 200$ \textmu m results in negligible change ($n'' \approx 7.7 \times 10^{-7}$), indicating that curvature-induced bending loss vanishes at the substrate leakage floor.
Consequently, for the experimental microresonator radius of $r = 227.5$ \textmu m, the bending loss contribution is minimal.
The residual substrate leakage of $n'' \approx 7.7 \times 10^{-7}$ corresponds to a propagation loss of $\alpha= 4\pi n'' / \lambda \cdot 10 \log_{10} e \approx 11.4$ dB m$^{-1}$.
This theoretical limit is over an order of magnitude lower than the experimentally measured background of $\sim$420 dB m$^{-1}$, confirming that material absorption---rather than structural leakage---is the dominant loss mechanism.

To examine the interaction between the optical mode and the cladding, we analyze the Poynting vector distributions.
Extended Data Fig. \ref{ExFig:3}b illustrates the normalized electric field magnitude $|E|$ of the fundamental transverse-electric (TE$_{00}$) mode at 81 THz.
Integration of the power flux reveals that the fraction of optical power propagating within the cladding rises from 21\% at 100 THz to 31\% at 81 THz.
This trend confirms reduced optical confinement at lower frequencies.
Given the onset of strong multi-phonon absorption in SiO$_2$ around 86 THz (3.5 \textmu m)~\cite{Soref:06b, Kitamura:07,Miller:17}, the increased mode overlap at 81 THz ($\sim$3.7 \textmu m) is consistent with the observed attenuation trend, indicating that interaction with the SiO$_2$ cladding is a substantial source of loss.

\vspace{0.3cm}
\noindent \textbf{Fog chamber}.
To simulate fog conditions, we use a 1.4-m-long environmental chamber equipped with an ultrasonic atomization unit. 
Water micro-droplets are introduced via a centrifugal fan, with scattering density regulated by a precision intake valve. 
A passive exhaust port enables natural atmospheric equilibration to maintain steady-state flow. 
Optical access is provided by broadband CaF$_2$ windows mounted at $\sim$10$^{\circ}$ angle relative to the optical axis to suppress etalon resonances. 
Resistive heating maintains window surface temperatures above the dew point to prevent condensation-induced scattering.

\medskip
\begin{footnotesize}

\noindent \textbf{Funding Information}: 
We acknowledge support from the National Key R\&D Program of China (Grant No. 2024YFA1409300),
National Natural Science Foundation of China (Grant No.12261131503, 12404417, 12404436, 62075233), 
Quantum Science and Technology--National Science and Technology Major Project (2023ZD0301500), 
Shenzhen-Hong Kong Cooperation Zone for Technology and Innovation (HZQB-KCZYB2020050), 
Shenzhen Science and Technology Program (Grant No. RCJC20231211090042078), 
Guangdong-Hong Kong Technology Cooperation Funding Scheme (Grant No. 2024A0505040008), 
the Key R\&D Plan of Shandong Province (Grant No. 2021ZDPT01), 
Natural Science Foundation of Shandong Province (ZR2021LLZ013, ZR2022LLZ009), 
and CAS Project for Young Scientists in Basic Research (YSBR-69). 

\noindent \textbf{Acknowledgments}: 
We thank Shui-Ming Hu for fruitful discussion on MIR light sources, 
Wei Sun, Jinbao Long and Shuyi Li for assistance in experiments,  
Sanli Huang, Jiahao Sun and Chen Shen for assistance in Si$_3$N$_4$ sample fabrication, 
and Jian-Wei Pan and Dapeng Yu for crucial support of this project. 

\noindent \textbf{Author contribution}: 
B. S., C. Z., Y. H., and Y.-H. L. built the setup and performed the experiment, with the assistance from X. B., H. G. and A. W.. 
M. Z., W. M. and X.-P. X. fabricated the CPLN devices, supervised by Q. Z..
Z. Z., Z. S. and J. L. fabricated the Si$_3$N$_4$ devices. 
B. S., C. Z., M. Z. and J. L. analyzed the data and prepared the manuscript with input from others. 
J. L. and Q. Z. managed the collaboration.
J. L. initiated and supervised the project. 

\noindent \textbf{Conflict of interest}:
B. S., Y.-H. L. and J. L. filed a patent application related to this work.  
J. L. and X. B. are developing heterogeneous silicon nitride integrated photonics technologies in Qaleido Photonics. 
Others declare no conflicts of interest.

\noindent \textbf{Data Availability Statement}: 
The code and data used to produce the plots within this work will be released on the repository \texttt{Zenodo} upon publication of this preprint.

\end{footnotesize}
\bibliographystyle{apsrev4-1}
\bibliography{bibliography}
\end{document}


\title{Supplementary Materials for: Metrology-grade mid-infrared spectroscopy for multi-dimensional perception}

\author{Baoqi Shi}
\affiliation{International Quantum Academy and Shenzhen Futian SUSTech Institute for Quantum Technology and Engineering, Shenzhen 518048, China}
\affiliation{Department of Optics and Optical Engineering, University of Science and Technology of China, Hefei 230026, China}

\author{Chenxi Zhang}
\affiliation{International Quantum Academy and Shenzhen Futian SUSTech Institute for Quantum Technology and Engineering, Shenzhen 518048, China}
\affiliation{College of Physics and Optoelectronic Engineering, Shenzhen University, Shenzhen 518060, China}

\author{Ming-Yang Zheng}
\affiliation{Jinan Institute of Quantum Technology and CAS Center for Excellence in Quantum Information and Quantum Physics, University of Science and Technology of China, Jinan 250101, China}
\affiliation{Hefei National Laboratory, University of Science and Technology of China, Hefei 230088, China}

\author{Yue Hu}
\affiliation{International Quantum Academy and Shenzhen Futian SUSTech Institute for Quantum Technology and Engineering, Shenzhen 518048, China}
\affiliation{Southern University of Science and Technology, Shenzhen 518055, China}

\author{Zeying Zhong}
\affiliation{International Quantum Academy and Shenzhen Futian SUSTech Institute for Quantum Technology and Engineering, Shenzhen 518048, China}
\affiliation{Southern University of Science and Technology, Shenzhen 518055, China}

\author{Zhenyuan Shang}
\affiliation{International Quantum Academy and Shenzhen Futian SUSTech Institute for Quantum Technology and Engineering, Shenzhen 518048, China}
\affiliation{Southern University of Science and Technology, Shenzhen 518055, China}

\author{Wenbo Ma}
\affiliation{Jinan Institute of Quantum Technology and CAS Center for Excellence in Quantum Information and Quantum Physics, University of Science and Technology of China, Jinan 250101, China}

\author{Xiu-Ping Xie}
\affiliation{Jinan Institute of Quantum Technology and CAS Center for Excellence in Quantum Information and Quantum Physics, University of Science and Technology of China, Jinan 250101, China}
\affiliation{Hefei National Laboratory, University of Science and Technology of China, Hefei 230088, China}

\author{Xue Bai}
\affiliation{International Quantum Academy and Shenzhen Futian SUSTech Institute for Quantum Technology and Engineering, Shenzhen 518048, China}
\affiliation{Qaleido Photonics, Shenzhen 518048, China}

\author{Yi-Han Luo}
\affiliation{International Quantum Academy and Shenzhen Futian SUSTech Institute for Quantum Technology and Engineering, Shenzhen 518048, China}

\author{Anting Wang}
\affiliation{Department of Optics and Optical Engineering, University of Science and Technology of China, Hefei 230026, China}

\author{Hairun Guo}
\affiliation{Key Laboratory of Specialty Fiber Optics and Optical Access Networks, Shanghai University, Shanghai 200444, China}

\author{Qiang Zhang}
\affiliation{Jinan Institute of Quantum Technology and CAS Center for Excellence in Quantum Information and Quantum Physics, University of Science and Technology of China, Jinan 250101, China}
\affiliation{Hefei National Laboratory, University of Science and Technology of China, Hefei 230088, China}
\affiliation{Hefei National Research Center for Physical Sciences at the Microscale and School of Physical Sciences, University of Science and Technology of China, Hefei 230026, China}
\affiliation{CAS Center for Excellence in Quantum Information and Quantum Physics, University of Science and Technology of China, Hefei 230026, China}

\author{Junqiu Liu}
\email[]{liujq@iqasz.cn}
\affiliation{International Quantum Academy and Shenzhen Futian SUSTech Institute for Quantum Technology and Engineering, Shenzhen 518048, China}
\affiliation{Hefei National Laboratory, University of Science and Technology of China, Hefei 230088, China}

\maketitle

\section{Spectroscopic system configuration and frequency metrology}
\vspace{0.5cm}

The experimental architecture, illustrated schematically in Supplementary Fig. \ref{SIfig:setup}, is designed for simultaneous multi-spectral operation across five distinct regions (\Rmnum{1}--\Rmnum{5}).
Regions \Rmnum{1} (1020--1098 nm) and \Rmnum{2} (1260--1640 nm) represent the fundamental outputs of the two near-infrared (NIR) external-cavity diode lasers (ECDLs, chirping lasers \#1 and \#2).
Chirping laser \#1 consists of a Toptica CTL laser capable of mode-hop-free tuning across 1020--1098 nm. 
Chirping laser \#2 comprises three Santec TSL lasers, which collectively cover the 1260--1360 nm, 1355--1485 nm, and 1480--1640 nm ranges.

These NIR seed lasers drive the generation of three derivative bands via nonlinear frequency conversion: 
visible output (Region \Rmnum{3}, 518--541 nm, via second harmonic generation(SHG)), mid-infrared (MIR) output (Region \Rmnum{4}, 3001--3711 nm, via difference frequency generation (DFG)), and near-visible output (Region \Rmnum{5}, 766--795 nm, via SHG).
Although the NIR seed lasers possess broad tuning ranges, the system's effective operational bandwidth is practically confined by the gain profiles of the Ytterbium-doped fiber amplifier (YDFA, 1035--1086 nm) and the Erbium-doped fiber amplifiers (EDFA, 1532--1590 nm).
Consequently, only spectral components residing within these gain windows achieve the power levels necessary to pump the chirped periodically poled lithium niobate (CPLN) waveguides, where chirped quasi-phase-matching (CQPM) conditions further select the efficient conversion bands.
This configuration specifically maps the amplified fundamental NIR scans to the target frequencies as follows: 
1036--1082 nm to 518--541 nm (Region \Rmnum{3}); 
1532--1590 nm to 766--795 nm (Region \Rmnum{5}); 
and combined pumps of 1035--1086 nm and 1536--1580 nm to 3001--3711 nm (Region \Rmnum{4}).

\begin{figure}[h!]
\centering
\renewcommand{\figurename}{Supplementary Figure}
\includegraphics{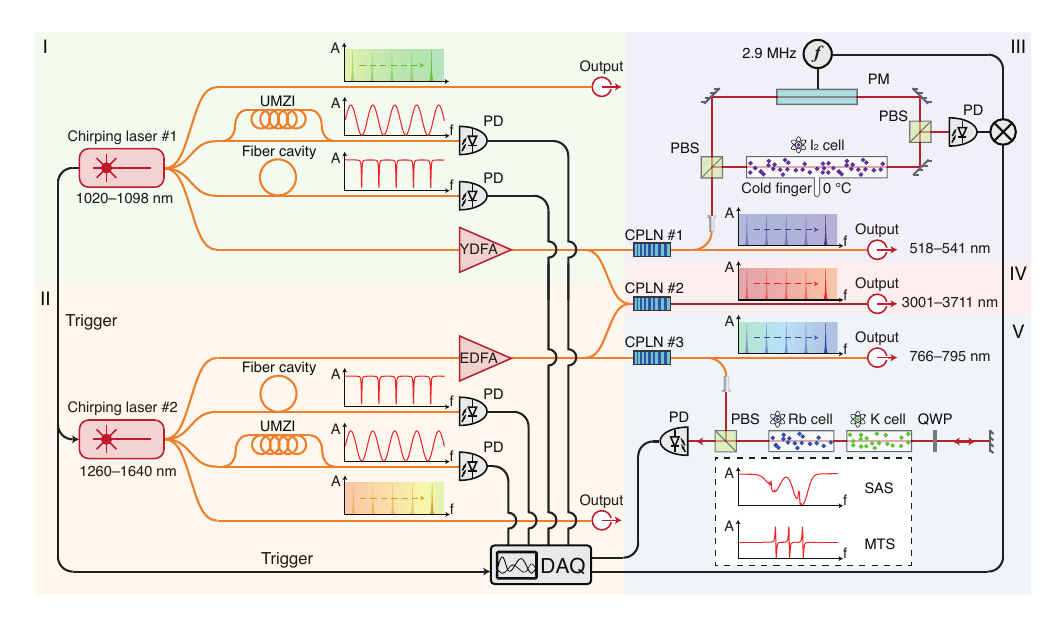}
\caption{
\textbf{Experimental setup}.
\textbf{a}.
Schematic of the spectroscopic system, functionally divided into five regions: 
Regions \Rmnum{1} and \Rmnum{2} track relative frequency calibration of the two NIR pump lasers via fiber cavities and UMZIs; 
Regions \Rmnum{3} and \Rmnum{5} establish absolute frequency referencing via frequency doubling to the visible/near-visible bands; 
and Region \Rmnum{4} generates the MIR DFG output. 
PD, photodetector; 
PBS, polarizing beam splitter;
QWP, quarter-wave plate;
DAQ, data acquisition;
PM, phase modulator.
Bottom-right inset illustrates the principle of saturated absorption spectroscopy (SAS) and modulation transfer spectroscopy (MTS).
}
\label{SIfig:setup}
\end{figure}

\subsection*{Relative frequency calibration (Regions \Rmnum{1} \& \Rmnum{2})}

To monitor the relative frequency evolution of the fundamental lasers during chirping, we employ fiber ring cavities with precisely characterized free spectral ranges (FSRs) as relative frequency rulers. 

\begin{figure}[h!]
\centering
\renewcommand{\figurename}{Supplementary Figure}
\includegraphics{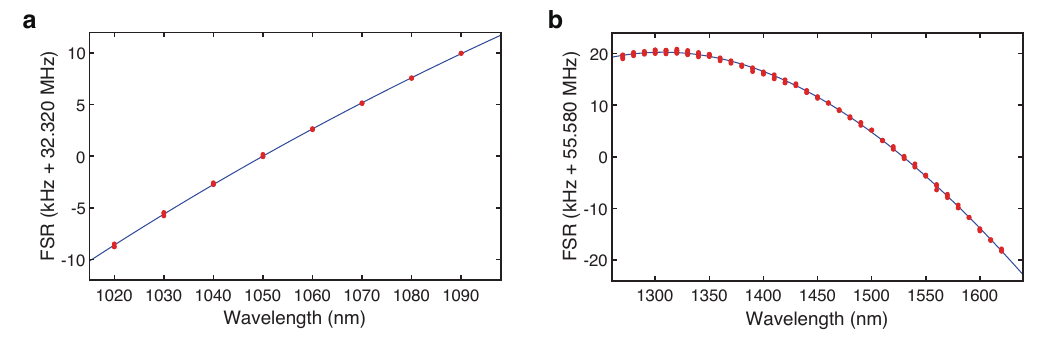}
\caption{
\textbf{Calibration of the fiber cavity}. 
\textbf{a}. 
Measured fiber cavity's free spectral range (FSR) variation from 1020 to 1098 nm.
The FSR curve is fitted with a quadratic polynomial formula.
\textbf{b}. 
Measured fiber cavity's FSR variation from 1260 to 1640 nm.
The FSR curve is fitted with a cubic polynomial formula.
}
\label{SIfig:fiberloop}
\end{figure}

For chirping laser \#1 in Region \Rmnum{1}, the fiber cavity FSR is calibrated using a sideband modulation method described in Ref.~\cite{Luo:24}. 
This technique utilizes three neighboring resonances to measure the local FSR at different wavelengths, accounting for dispersion-induced variations over the measurement bandwidth. 
We measure the FSR from 1020 to 1098 nm and fit the data using a quadratic polynomial (Supplementary Fig. \ref{SIfig:fiberloop}a), enabling dispersion calibration up to the second order. 

Similarly, for chirping laser \#2 in Region \Rmnum{2}, a second fiber ring cavity is employed. 
Its FSR is measured across 1260 to 1640 nm and fitted with a cubic polynomial to calibrate dispersion up to the third order (Supplementary Fig. \ref{SIfig:fiberloop}). 

By recording and counting the number of fiber resonances passed by the chirping laser, the relative frequency excursion of the chirping laser is calculated. 
While the fiber resonances provide discrete frequency markers, sub-FSR instantaneous frequencies are resolved by interpolating interference fringes from an unbalanced Mach-Zehnder interferometer (UMZI) \cite{Ahn:05}.
Using the Hilbert transform \cite{Marple:99} to extract the instantaneous phase $\phi(t)$, we calculate the phase difference $\Delta \phi$ between adjacent sampling points. 
The corresponding frequency excursion $\Delta f$ is then derived as:
\begin{equation}
  \Delta f = \frac{\text{FSR}}{\phi_\text{FSR}} \cdot \Delta \phi,
\label{Eq.UMZI}
\end{equation}
where $\phi_\text{FSR}$ is the phase change corresponding to one cavity FSR.

\subsection*{Visible band and absolute calibration (Region \Rmnum{3})}

In Region \Rmnum{3}, the output of chirping laser \#1 is amplified by a YDFA and undergoes SHG in a CPLN waveguide (CPLN\#1) to produce tunable output from 518 to 541 nm. 
To enable broadband SHG phase matching, the poling period along CPLN\#1 is linearly chirped from 6.3 to 7.3 \textmu m.
The tuning curve is shown in Supplementary Fig. \ref{SIfig:532}a, with the second-harmonic (SH) power reaching 45.9 mW under a 5.0-W pump (Supplementary Fig. \ref{SIfig:532}b). 

Absolute frequency calibration in this region is achieved via modulation transfer spectroscopy (MTS) \cite{Shirley:82, Cheng:19, Shi:25} of molecular iodine ($\text{I}_2$). 
As shown in Supplementary Fig. \ref{SIfig:setup}, the setup utilizes a 50-cm-long $\text{I}_2$ vapor cell with its cold finger stabilized at $0~^\circ\text{C}$ via a thermoelectric cooler (TEC) to suppress collision broadening. 
A 5-mW pump beam, modulated at 2.9~MHz to generate sidebands, interacts with a 0.5-mW probe beam. 
When the laser resonates with a hyperfine transition, four-wave mixing transfers the sidebands to the probe, which is subsequently demodulated to extract the MTS signal. 
We conduct MTS measurements over the 518--541~nm band (subset 530--540~nm shown in Supplementary Fig. \ref{SIfig:532}c).

To establish absolute frequency references, we utilize the hyperfine transitions of the I$_2$ R(56)32-0, R(53)31-0, and P(28)30-0 lines (zoomed view in Supplementary Fig. \ref{SIfig:532}d). 
The frequencies of the hyperfine transitions used for absolute referencing are listed in Supplementary Table~\ref{tab:iodine}.
The absolute frequency axis is anchored to the I$_2$ R(56)32-0 line ($f_{\text{ref,1}} = 563.2600$ THz)~\cite{Arie:93}. 
We use the hyperfine transitions of the I$_2$ P(28)30-0 line ($f_{\text{ref,2}} = 560.1518$~THz) to validate the relative frequency calibration derived from the fiber cavity. 
We observe a root-mean-square (RMS) deviation of $\Delta_1=-35.8$ MHz between our measured values and those in Ref.~\cite{Shie:13}, which we attribute to residual dispersion in the fiber cavity's FSR. 
Consequently, we apply a correction to the SH laser frequency using:
\begin{equation}
f'_{\text{SH1}} = \frac{f_{\text{SH1}} - f_{\text{ref,1}}}{1 + \epsilon_1} + f_{\text{ref,1}}
\label{Eq.Correct}
\end{equation}
where $\epsilon_1 = \Delta_1/(f_{\text{ref,1}} - f_{\text{ref,2}})$. 
Applying Eq. \ref{Eq.Correct} to the measurement of the I$_2$ R(53)31-0 transitions ($f_{\text{ref,3}} = 561.4205$ THz) yields a residual RMS deviation of 11.7 MHz relative to literature values \cite{Cheng:19}. 
This determines the SH laser accuracy as $\delta_\text{SH1}\approx 11.7$ MHz, corresponding to a fundamental frequency accuracy for chirped laser \#1 of $\delta_{\text{laser1}} = \delta_{\text{SH1}}/2 \approx 5.9 $ MHz.

\begin{figure}[h!]
\centering
\renewcommand{\figurename}{Supplementary Figure}
\includegraphics{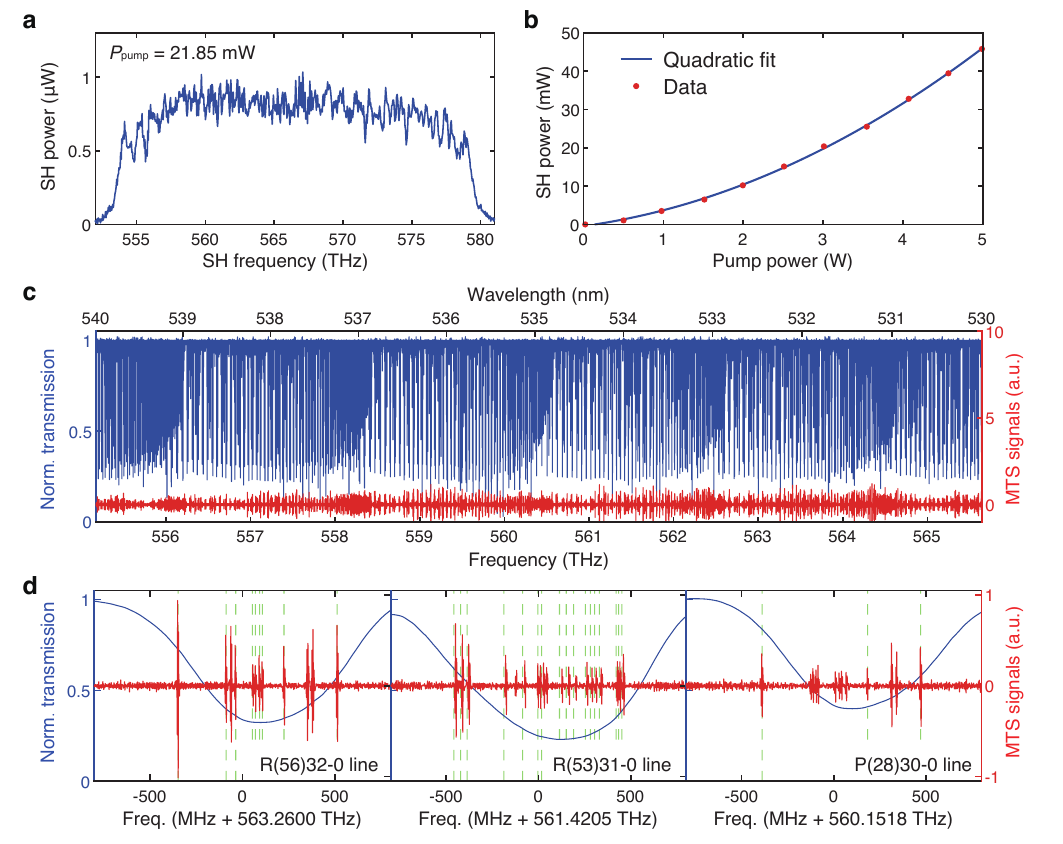}
\caption{
\textbf{Broadband SHG performance in the visible band and absolute frequency references}. 
\textbf{a}.
Bandwidth of the SH laser generated by the CPLN\#1 waveguides.
$P_\mathrm{pump}=21.85$ mW.
\textbf{b}.
SH power generated by the CPLN\#1 waveguide for 518--541 nm SHG, with different pump power $P_\mathrm{pump}$ at 1064 nm. 
\textbf{c}.
Absorption spectrum and modulation transfer spectrum of I$_2$.
\textbf{d}.
The measured R(56)32-0, R(53)31-0, and P(28)30-0 lines of I$_2$. 
The MTS signals clearly resolve the hyperfine transitions associated with these lines, which are employed as absolute frequency references for chirping laser \#1.
}
\label{SIfig:532}
\end{figure}
\begin{figure}[b!]
\centering
\renewcommand{\figurename}{Supplementary Figure}
\includegraphics{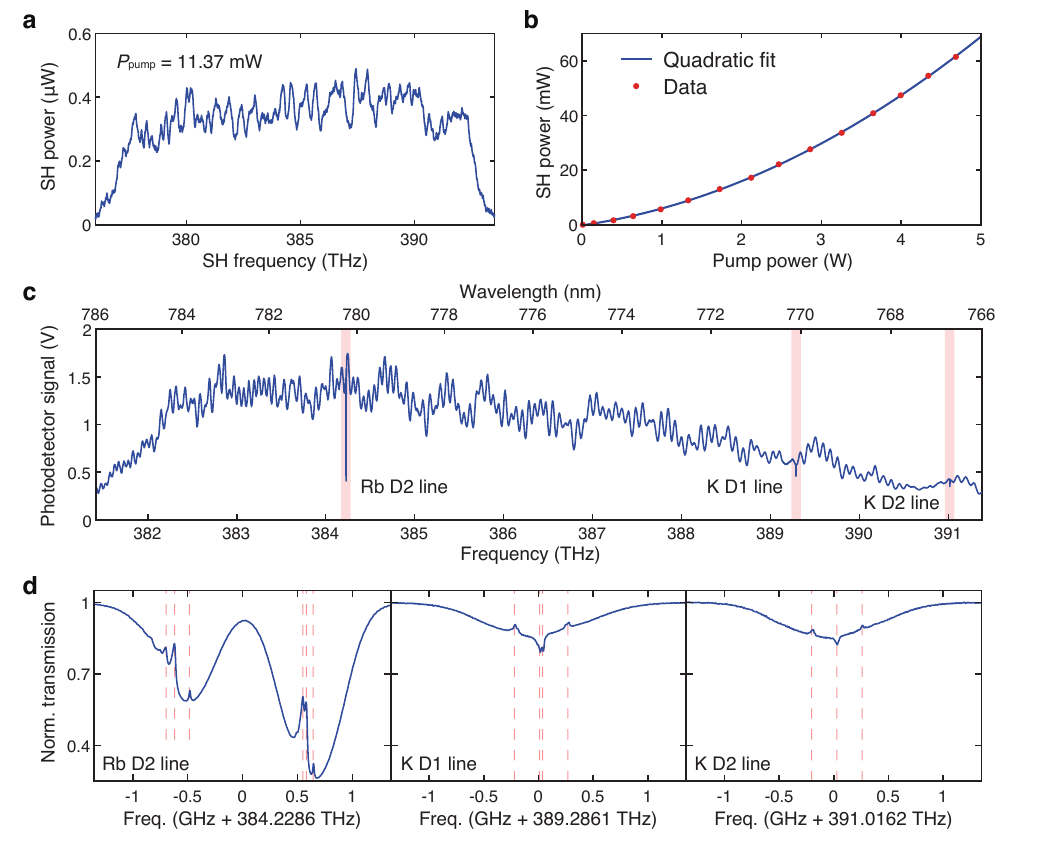}
\caption{
\textbf{Broadband SHG performance in the near-visible band and absolute frequency references}. 
\textbf{a}.
Bandwidth of the SH laser generated by the CPLN\#3 waveguides.
$P_\mathrm{pump}=11.37$ mW.
\textbf{b}.
SH power generated by the CPLN\#3 waveguide for 766--795 nm SHG, with different pump power $P_\mathrm{pump}$ at 1560 nm. 
\textbf{c}.
Saturated absorption spectrum of Rb and K atoms.
\textbf{d}.
Saturated absorption spectrum of Rb D2, K D1, and K D2 lines.
The thirteen peaks used for the absolute frequency reference are marked with red dashed lines.
}
\label{SIfig:780}
\end{figure}

\subsection*{Near-visible band and absolute frequency calibration (Region \Rmnum{5})}

In Region \Rmnum{5}, chirping laser \#2 is amplified by an EDFA and frequency-doubled in CPLN\#3 to generate 766--795~nm light. 
To enable broadband SHG phase matching, the poling period along CPLN\#3 is linearly chirped from 18.1 to 19.9 \textmu m.
The tuning curve is shown in Supplementary Fig. \ref{SIfig:780}a, with the SH power reaching 68.6 mW under a 5.0-W pump (Supplementary Fig. \ref{SIfig:780}b). 

Absolute frequency calibration is performed using saturated absorption spectroscopy (SAS) of Rubidium (Rb) and Potassium (K). 
The setup (Supplementary Fig. \ref{SIfig:setup}) employs two cascaded 10-cm vapor cells containing Rb and K, respectively. 
Two counter-propagating SH beams overlap within the cells; 
the strong pump beam creates a transparency window for the probe beam when resonant with a hyperfine transition, resulting in narrow peaks within the Doppler-broadened profile. 

The photodetected light signal from the vapor cells is shown in Supplementary Fig. \ref{SIfig:780}c, where Rb D2 line, K D1 and D2 lines are resolved.
Supplementary Fig. \ref{SIfig:780}d shows the zoom-in profile of Rb D2, K D1, and K D2 lines.  
The specific hyperfine transitions identified and their corresponding literature frequencies are detailed in Supplementary Table~\ref{tab:RbK}. 

The absolute frequency axis is anchored to the Rb D2 line ($f_{\text{ref,4}} = 384.2286$ THz)~\cite{Steck:23}.
The hyperfine transitions associated with the K D2 line ($f_{\text{ref,5}} = 391.0165$ THz)~\cite{Tiecke:19} are used to examine the accuracy of our relative frequency calibration with the fiber cavity.
An RMS deviation of $\Delta_2=28.4$ MHz is found between our measured values and the values in Ref. \cite{Tiecke:19}.
We further correct the SH laser's frequency as:
\begin{equation}
f'_{\text{SH2}} = \frac{f_{\text{SH2}} - f_{\text{ref,4}}}{1 + \epsilon_2} + f_{\text{ref,4}}
\label{Eq.Correct2}
\end{equation}
where $\epsilon_2 = \Delta_2/(f_{\text{ref,4}} - f_{\text{ref,5}})$.
Applying Eq. \ref{Eq.Correct2} to the measurement of the K D1 line ($f_{\text{ref,6}} = 389.2861$ THz) yields a residual RMS deviation of 8.2 MHz relative to literature values \cite{Tiecke:19}.
This determines the SH laser accuracy as $\delta_\text{SH2}\approx 8.2$ MHz, corresponding to a fundamental accuracy for chirping laser \#2 of $\delta_{\text{laser2}} = \delta_{\text{SH2}}/2 \approx 4.1$ MHz.

\subsection*{Mid-IR band (Region \Rmnum{4})}

In Region \Rmnum{4}, mid-infrared radiation (3001--3711 nm) is generated via DFG. 
The amplified outputs of chirping laser \#1 and \#2 are combined using a high-power fiber wavelength division multiplexer (WDM) and coupled into CPLN\#2. 
The waveguide output is collimated into free space using a Calcium Fluoride ($\text{CaF}_2$) lens, and the residual pump beams are filtered using a Germanium (Ge) window.

\begin{table}[h!]
\centering
\caption{\textbf{Reference frequencies of $\text{I}_2$ hyperfine transitions used for calibration in Region \Rmnum{3}.}}
\label{tab:iodine}
\renewcommand{\arraystretch}{1.2}
\begin{tabular}{p{2.5cm} p{1.5cm} p{10cm}}
\toprule
\textbf{Spectral line} & \textbf{Ref.} & \textbf{Frequencies (THz)} \\ 
\addlinespace[0.5 em]
\midrule
R(56)32-0 & \cite{Arie:93} & 563.259651971, 563.259911669, 563.259963337, 563.260053449, 563.260068965, 563.260091597, 563.260107314, 563.260223513, 563.260509925 \\ 
\addlinespace[0.5 em]
\midrule
R(53)31-0 & \cite{Cheng:19} & 561.420041099, 561.420078214, 561.420113401, 561.420312353, 561.420413980, 561.420496396, 561.420517033, 561.420613626, 561.420650083, 561.420689990, 561.420753971, 561.420780437, 561.420803713, 561.420829676, 561.420920348, 561.420934177, 561.420951343 \\ 
\addlinespace[0.5 em]
\midrule
P(28)30-0 & \cite{Shie:13} &  560.151411260, 560.151982649, 560.152269813 \\ 
\bottomrule
\end{tabular}
\end{table}

\begin{table}[h!]
\centering
\caption{\textbf{Reference frequencies of Rb and K saturated absorption peaks used for calibration in Region \Rmnum{5}.} $CO(i,j)$ and $CO'(i,j)$ denote ground-state and excited-state crossovers, respectively.}
\label{tab:RbK}
\renewcommand{\arraystretch}{1.3}
\begin{tabular}{p{2.5cm} p{1.5cm} p{5cm} p{5cm}}
\toprule
\textbf{Line} & \textbf{Ref.} & \textbf{Assignment} & \textbf{Frequency (THz)} \\ 
\addlinespace[0.5 em]
\midrule
$^{87}$Rb D2 & \cite{Steck:23} & $F=2\rightarrow CO'(1,3)$ & 384.227903407 \\
             &                 & $F=2\rightarrow CO'(2,3)$ & 384.227981877 \\
             &                 & $F=2\rightarrow F'=3$      & 384.228115203 \\
\addlinespace[0.5 em]
\midrule
$^{85}$Rb D2 & \cite{Steck:23} & $F=3\rightarrow CO'(2,4)$ & 384.229149669 \\
             &                 & $F=3\rightarrow CO'(3,4)$ & 384.229181369 \\
             &                 & $F=3\rightarrow F'=4$      & 384.229241689 \\
\addlinespace[0.5 em]
\midrule
K D1         & \cite{Tiecke:19}& $F=2\rightarrow CO'(1,2)$     & 389.2858787 \\
             &                 & $CO(1,2)\rightarrow CO'(1,2)$ & 389.2861095 \\
             &                 & $CO(1,2)\rightarrow F'=2$     & 389.2861373 \\
             &                 & $F=1\rightarrow F'=2$         & 389.2863681 \\
\addlinespace[0.5 em]
\midrule
K D2         & \cite{Tiecke:19}& $F=2\rightarrow F'$         & 391.0159969 \\
             &                 & $CO(1,2)\rightarrow F'$     & 391.0162278 \\
             &                 & $F=1\rightarrow F'$         & 391.0164586 \\
\bottomrule
\end{tabular}
\end{table}
\clearpage

\section{Design and fabrication of CPLN waveguides}
\vspace{0.5cm}
\subsection*{Device design and theoretical framework}
\begin{figure}[h!]
\centering
\renewcommand{\figurename}{Supplementary Figure}
\includegraphics{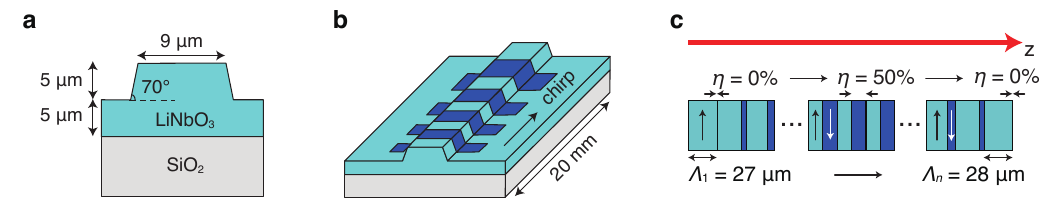}
\caption{
\textbf{Design and geometry of the CPLN waveguide.}
\textbf{a}.
Schematic illustration of the ridge waveguide geometry. 
LiNbO$_3$, lithium niobate.
SiO$_2$, silicon dioxide.
\textbf{b}.
Longitudinal profile showing the poling period $\Lambda$ and duty cycle $\eta$ of the CPLN waveguide. 
}
\label{SIfig:CPLNdesign}
\end{figure}

In this work, we design and fabricate CPLN waveguides to generate broadband MIR laser spanning 3001--3711~nm via DFG. 
Supplementary Fig.~\ref{SIfig:CPLNdesign}a and b illustrate the cross-sectional and longitudinal geometry of the device.
The lithium niobate (LiNbO$_3$) ridge waveguide features a top width of 9~\textmu m, an etching depth of 5~\textmu m, a total thickness of 10~\textmu m, and a sidewall angle of 70$^{\circ}$. 
The total device length is 20~mm.
To satisfy the CQPM condition over a broad spectral range, the poling period ($\Lambda$) is linearly chirped from 27~\textmu m to 28~\textmu m along the propagation direction.

In the low-conversion-efficiency regime (undepleted pump approximation) and assuming negligible group-velocity mismatch, the theoretical DFG spectral response can be described by the Fourier transform of the nonlinear grating profile:
\begin{equation}
\left|\widehat{D}(\omega)\right|^2= \left|\frac{2 \pi}{\lambda_{\mathrm{signal}} n_{\mathrm{idler}}} \int_{-\infty}^{\infty}\mathrm{d} z~\tilde{d}(z) \exp \left(-i \Delta k^{\prime} z\right) \right|^2,
\label{Eq.D}
\end{equation}
Here, the term $\Delta k^{\prime} = k_{\mathrm{pump}} - k_{\mathrm{signal}} - k_{\mathrm{idler}}$ represents the wave-vector mismatch attributed to material dispersion, where $k$ is the wave vector at the corresponding optical frequency.
The subscripts `pump', `signal', and `idler' refer to the input fields (from chirping laser \#1 and chirping laser \#2) and the generated MIR output, respectively. 
$n_{\mathrm{idler}}$ denotes the effective refractive index at the idler wavelength.
The complex function $\tilde{d}(z)$ describes the effective nonlinear optical coefficient distribution along the propagation axis $z$:
\begin{equation}
  \tilde{d}(z) = \mathrm{rect}(z/L) \, \frac{2}{\pi}\sin[\pi \eta(z)] \exp\left[-i \int_0^z K_\mathrm{CQPM}(z')\mathrm{d}z'\right],
  \label{Eq.d}
\end{equation}
where $\eta(z)$ is the poling duty cycle, $L$ is the interaction length, and $K_{\mathrm{CQPM}}(z) = 2\pi/\Lambda(z)$ corresponds to the local grating vector.
As implied by Eq.~\ref{Eq.D}, the DFG tuning curve maps directly to the spatial frequency spectrum of the grating structure~\cite{Imeshev:00}.
The quasi-phase-matching structure is characterized by a spatially dependent poling period $\Lambda(z)$ and duty cycle $\eta(z)$, as detailed in Supplementary Fig.~\ref{SIfig:CPLNdesign}c. 
Here, $\eta(z)$ represents the fraction of the domain inverted within a single local period.

\subsection*{Duty cycle apodization for spectral flatness}
\begin{figure}[h!]
\centering
\renewcommand{\figurename}{Supplementary Figure}
\includegraphics{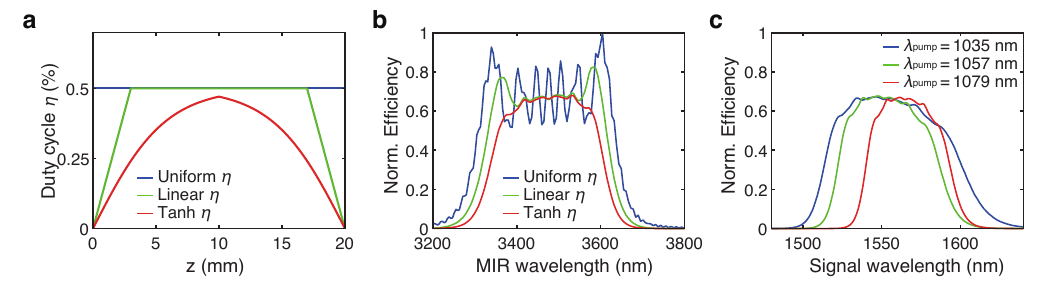}
\caption{
\textbf{Optimization of the poling duty cycle profile for broadband spectral flatness.}
\textbf{a}. 
Spatial distribution of the duty cycle $\eta$ along the propagation direction. Three profiles are compared: a uniform distribution (blue), a linear apodization scheme (green), and a hyperbolic tangent (tanh) apodization profile with a shape parameter $\beta=7$ (red).
\textbf{b}. 
Theoretical DFG tuning curves corresponding to the three waveguide designs.
\textbf{c}. 
Theoretical DFG tuning curves corresponding to different pump wavelength $\lambda_{\mathrm{pump}}$.
}
\label{SIfig:CPLNeta}
\end{figure}

To achieve a broadband MIR laser with a spectrally flat profile, optimizing the spectral response of the DFG process is critical. 
We evaluate a CPLN waveguide with a uniform duty cycle ($\eta=50\%$) along the entire propagation length (Supplementary Fig.~\ref{SIfig:CPLNeta}a, blue trace). 
By simulating the DFG process with a fixed pump at 1079~nm and a signal sweep from 1500 to 1650~nm, the theoretical tuning curve predicts an MIR spectral coverage spanning 3309--3632~nm (Supplementary Fig.~\ref{SIfig:CPLNeta}b). 
However, the abrupt truncation of the nonlinear interaction domain introduces significant passband ripples, known as Gibbs oscillations.

To suppress these efficiency oscillations, we implement spatial apodization of the poling duty cycle $\eta$.
A linear apodization scheme, where $\eta$ ramps linearly from 0\% to 50\% over the initial 3~mm and tapers symmetrically over the final 3~mm (Supplementary Fig.~\ref{SIfig:CPLNeta}a, green trace).
The theoretical tuning curve shows the ripples are mitigated to some extent (Supplementary Fig.~\ref{SIfig:CPLNeta}b). 
Nevertheless, noticeable fluctuations persist near the band edges of the tuning curve.
Consequently, we adopt a more robust hyperbolic tangent (tanh) apodization profile (Supplementary Fig.~\ref{SIfig:CPLNeta}a, red trace), defined as:
\begin{equation}
\eta(z) =
\begin{cases} 
0.5 \tanh\left(\frac{\beta z}{L}\right) & 0 < z \le L/2 \\
0.5 \tanh\left[\frac{\beta (L-z)}{L}\right] & L/2 < z \le L 
\end{cases}
\end{equation}
where $L$ is the waveguide length and $\beta$ is the apodization parameter~\cite{Umeki:09}. 
By setting $\beta = 7$, this profile effectively eliminates the hard truncation of the interaction domain, thereby suppressing Gibbs oscillations while preserving the bandwidth (3330--3615 nm). 
The resulting tuning curve exhibits a desirable flat-top spectral profile, as shown in Supplementary Fig.~\ref{SIfig:CPLNeta}b.

While apodization ensures a spectrally flat response, the broadband capability inherently stems from the chirped grating structure. 
This design provides a continuous distribution of reciprocal lattice vectors ($K_{\mathrm{CQPM}}$), allowing the CQPM condition to be satisfied over a wide range of pump and signal wavelengths. 
Physically, shifting the pump or signal wavelength merely relocates the localized interaction region along the waveguide's longitudinal axis without sacrificing conversion efficiency.

Capitalizing on this broad CQPM acceptance, we implement a pump tuning strategy to circumvent the bandwidth limitations of a fixed pump wavelength. 
As illustrated in Supplementary Fig.~\ref{SIfig:CPLNeta}c, by switching $\lambda_{\mathrm{pump}}$ between 1035~nm, 1057~nm, and 1079~nm, the phase-matching condition is consistently satisfied when tuning the signal wavelength across 1536--1580~nm. 
Thus, we can generate a broadband mid-infrared laser spectrum by tuning the pump and signal lasers together.
This dual-tuning scheme yields continuous coverage from 3001~nm to 3711~nm, as demonstrated in the main text, showing that synchronizing tunable lasers operating within the YDFA (pump) and EDFA (signal) gain bands effectively accesses the entire molecular functional group region.

\subsection*{Fabrication process of CPLN waveguide}
The fabrication workflow of the CPLN waveguide is illustrated in Supplementary Fig. \ref{SIfig:CPLN}. 
The device is fabricated on a commercially available lithium-niobate-on-insulator (LNOI) wafer (NANOLN Inc.), consisting of a 10-\textmu m-thick z-cut MgO-doped LiNbO$_3$ film bonded to a 2-\textmu m-thick thermal SiO$_2$ layer on a silicon substrate. 
First, Aaluminum (Al) electrodes are patterned on the LiNbO$_3$ surface via electron-beam evaporation followed by standard UV lithography and wet etching. 
Periodic poling is then achieved by applying high-voltage pulses across the electrodes. 
Subsequently, the aluminum electrodes are removed, and a chromium (Cr) hard mask is sputtered and patterned on the film. 
Finally, the ridge waveguide geometry is transferred to the LiNbO$_3$ layer using inductively coupled plasma (ICP) etching.

\begin{figure}[h!]
\centering
\renewcommand{\figurename}{Supplementary Figure}
\includegraphics[width=\linewidth]{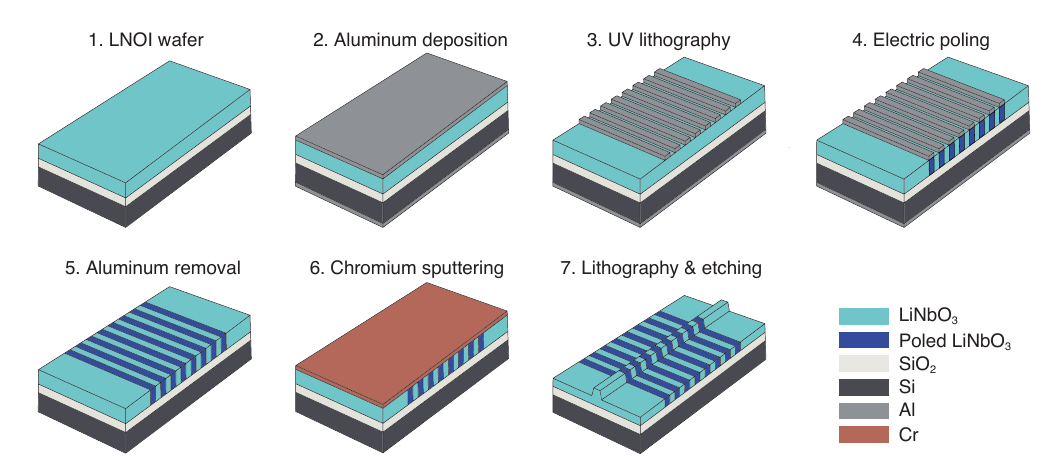}
\caption{
\textbf{Fabrication process flow of CPLN waveguides}. 
LNOI, lithium-niobate-on-insulator.
Si, silicon.
Al, aluminum.
Cr, chromium.
}
\label{SIfig:CPLN}
\end{figure}
\clearpage

\section{The MIR laser's dynamic linewidth}
\vspace{0.5cm}

\begin{figure}[h!]
\centering
\renewcommand{\figurename}{Supplementary Figure}
\includegraphics[width=\linewidth]{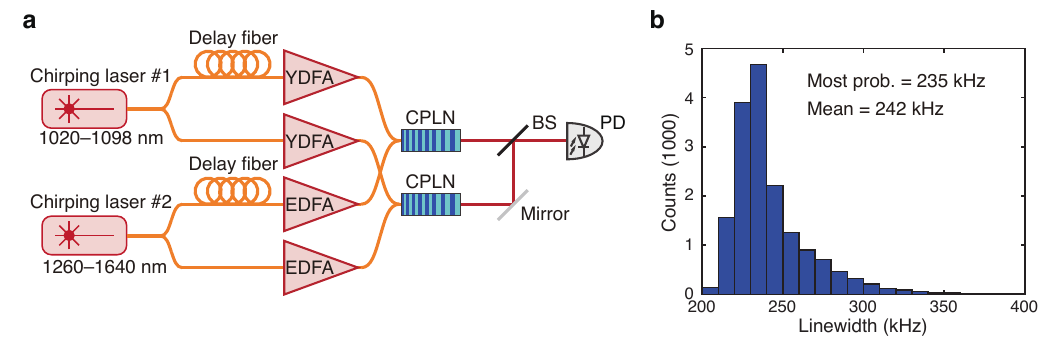}
\caption{
\textbf{a}. 
Schematic of the self-delayed heterodyne measurement. 
BS, beam splitter.
\textbf{b}. 
Histogram of measured MIR chirping laser's dynamic linewidth at 3.56 \textmu m. 
Here 16,775 segments of beatnote signals are taken and processed with STFT to extract the laser linewidth. 
}
\label{SIfig:linewidth}
\end{figure}

Supplementary Fig.~\ref{SIfig:linewidth}a illustrates the self-delayed heterodyne setup employed to characterize the dynamic linewidth of the MIR laser. 
To ensure signal decoherence, chirping laser \#1 is split into two paths. 
One path propagates through a 20,408-meter-long optical fiber, introducing a time delay that exceeds the laser's coherence time. 
Both paths are subsequently amplified by YDFAs. 
Similarly, chirping laser \#2 is split into two paths, with one path delayed by a 20,216-meter-long fiber;
both are amplified by EDFAs. 
The non-delayed pumps from laser \#1 and laser \#2 undergo DFG in a CPLN waveguide to generate the first MIR beam. 
Concurrently, the delayed pumps generate a second MIR beam in a separate CPLN waveguide. 
Consequently, the two generated MIR beams are mutually incoherent. 
These beams are spatially combined, and their interference beat signal is recorded by a photodetector (PD).

During the measurement centered at 3.56 \textmu m, the laser is chirped at a rate of approximately 1.4~THz/s. 
The temporal beat signal is recorded and transformed into the frequency domain using a short-time Fourier transform (STFT) with a window length of 100 \textmu s. 
We analyzed the linewidths of 16,775 beatnote signals, and the resulting histogram is presented in Supplementary Fig.~\ref{SIfig:linewidth}b. 
The statistical linewidth corresponding to the 100-\textmu s window is determined to be 242~kHz, with a most probable value of 235~kHz. 
This measured linewidth accounts for the laser's intrinsic linewidth, chirp nonlinearity, and fluctuations in the fiber delay lines. 
Based on these results, we determine the frequency resolution of our spectroscopic system to be 242~kHz.

\vspace{0.5cm}
\section{Mid-infrared microresonator characterization}
\vspace{0.5cm}
Characterization of the Si$_3$N$_4$ microresonators is performed across the 3001--3711 nm band (81--100 THz). 
The devices feature a uniform waveguide width of 3 \textmu m, with bus-to-microresonator coupling gaps of 300, 350, 400, and 450 nm. 
To extract the loss parameters, the measured resonances are fitted to a Lorentzian model \cite{Aspelmeyer:14, Pfeiffer:18}:
\begin{equation}
    T(\omega) = 1 - \frac{4\kappa_\text{ex}\kappa_0}{\Delta\omega^2 + (\kappa_\text{ex}+\kappa_0)^2}
\label{Eq.Res}
\end{equation}
where $\Delta\omega/2\pi$ is the laser detuning, 
$\kappa_0/2\pi$ is the intrinsic loss, 
and $\kappa_\text{ex}/2\pi$ is the external coupling strength. 
The results are summarized in Supplementary Fig. \ref{SIfig:chip}a.

Within the 93--95 THz window, the extracted $\kappa_0$ and $\kappa_{\mathrm{ex}}$ for the as-deposited resonators exhibit a localized discontinuity. 
This spectral region corresponds to a crossover point where the onset of strong material absorption drives the system from the over-coupled to the under-coupled regime. 
While the transmission profile is mathematically invariant under the exchange of $\kappa_0$ and $\kappa_{\mathrm{ex}}$ (Eq. \ref{Eq.Res}), we resolve this ambiguity by verifying the exponential dependence of $\kappa_{\mathrm{ex}}$ on the gap size (Supplementary Fig. \ref{SIfig:chip}a). 
The observed numerical artifacts near the crossover arise as these two fitted parameters become quantitatively indistinguishable ($\kappa_0 \approx \kappa_{\mathrm{ex}}$). 
In this limit, experimental noise factors---such as photodetector dark current and background fluctuations---obscure the difference between them, inducing numerical deviations that manifest as the observed discontinuity.

The corresponding linear propagation losses, $\alpha$, are summarized in Supplementary Fig. \ref{SIfig:chip}b. 
The as-deposited Si$_3$N$_4$ microresonators display strong absorption features approaching 100 THz, attributed to N-H bond stretching, with peak losses ranging from $\sim$1900 to 2650 dB m$^{-1}$ across the measured devices. 
Crucially, high-temperature thermal annealing suppresses these losses by two orders of magnitude, reducing the propagation loss to $\sim$30 dB m$^{-1}$ at 100 THz.

\begin{figure}[h!]
\centering
\renewcommand{\figurename}{Supplementary Figure}
\includegraphics[width=\linewidth]{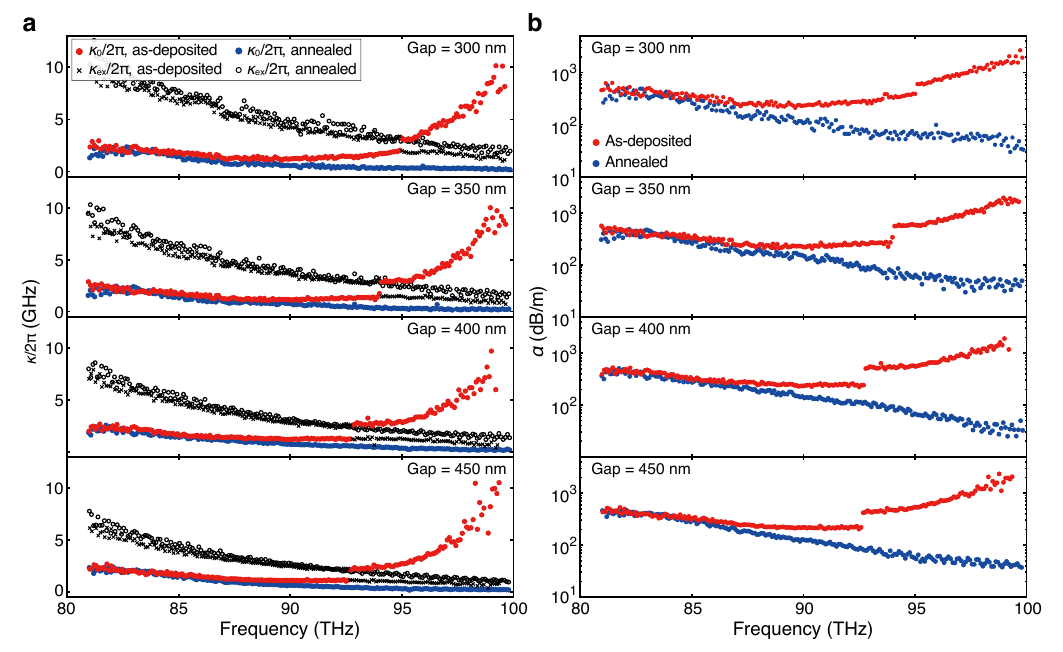}
\caption{
\textbf{Characterization of Si$_3$N$_4$ microresonators}. 
\textbf{a}. 
Extracted intrinsic loss ($\kappa_0/2\pi$) and external coupling strength ($\kappa_{\mathrm{ex}}/2\pi$) as a function of frequency for waveguide-to-microresonator gaps of 300, 350, 400, and 450 nm.
The gap-dependent trend of $\kappa_{\mathrm{ex}}$ allows for the identification of coupling regimes. 
\textbf{b}. 
Frequency-dependent linear propagation loss ($\alpha$). 
The as-deposited devices exhibit strong absorption peaks near 100 THz attributed to N-H bond stretching. High-temperature annealing suppresses these losses by approximately two orders of magnitude, reducing the background loss to $\sim$30 dB m$^{-1}$.
}
\label{SIfig:chip}
\end{figure}
\clearpage

\section{Fabrication process of silicon nitride integrated waveguide}
\vspace{0.5cm}
The silicon nitride (Si$_3$N$_4$) integrated waveguides are fabricated on 150-mm (6-inch) silicon wafers employing an optimized deep-ultraviolet (DUV) subtractive process \cite{Ye:23}, as depicted in Supplementary Fig. \ref{SIfig:SiN}.
Initially, a 1050-nm-thick Si$_3$N$_4$ film is deposited via low-pressure chemical vapor deposition (LPCVD) onto a silicon substrate featuring a thermally grown wet silicon dioxide (SiO$_2$) buffer layer.
To mitigate the high intrinsic tensile stress ($\sim$1.1--1.4 GPa) characteristic of LPCVD Si$_3$N$_4$ films, we employ a multi-step deposition method interspersed with thermal cycling \cite{Gondarenko:09}. 
This approach effectively relieves film stress, yielding crack-free wafers.
Subsequently, a SiO$_2$ hardmask layer is deposited via LPCVD.
Device patterns are defined using DUV stepper lithography and transferred into both the SiO$_2$ hardmask and the underlying Si$_3$N$_4$ layer through dry etching.
An optimized CHF$_3$/O$_2$ gas chemistry is utilized during etching to ensure ultra-smooth, vertical sidewalls, a critical factor for minimizing optical scattering losses.
After removing the photoresist, the wafers undergo thermal annealing over 1200 $^\circ$C in a nitrogen atmosphere.
This high-temperature annealing step drives out residual hydrogen impurities, thereby significantly suppressing optical absorption losses associated with N-H bonds.
Finally, a 3-\textmu m-thick SiO$_2$ top cladding is deposited, followed by a second annealing step at 1200 $^\circ$C. 
The chip facets and dimensions are defined via UV photolithography and deep dry etching, and the individual chips are singulated through backside grinding and dicing.

\begin{figure}[h!]
\centering
\renewcommand{\figurename}{Supplementary Figure}
\includegraphics[width=\linewidth]{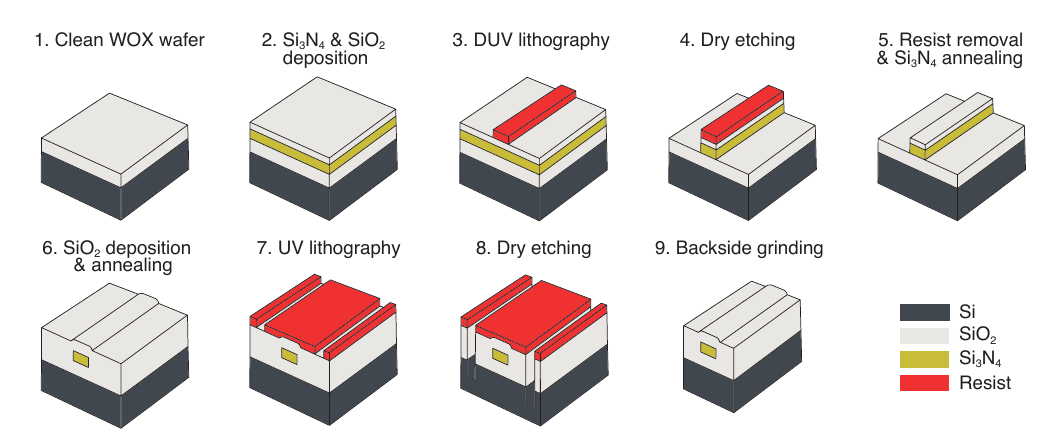}
\caption{
\textbf{The DUV subtractive process flow of 6-inch-wafer Si$_3$N$_4$ foundry fabrication}. 
WOX, thermal wet oxide (SiO$_2$).
}
\label{SIfig:SiN}
\end{figure}
\clearpage

\section{Linearization of chirp nonlinearity for FMCW LiDAR}
\vspace{0.5cm}
In practical frequency-modulated continuous-wave (FMCW) systems, deviations from perfect linearity in the frequency chirp are unavoidable \cite{Shi:24}. 
The instantaneous laser frequency, $f_\text{L}(t)$, deviates from the ideal linear ramp ($f_0 + \gamma t$) due to realistic device constraints, and is described by:
$$f_\text{L}(t) = f_0 + \gamma t + \delta \nu(t),$$
where $f_0$ is the start frequency, $\gamma$ is the nominal chirp rate, and $\delta \nu(t)$ represents the time-dependent frequency nonlinearity.
Fundamentally, the instantaneous optical phase $\phi(t)$ is related to the frequency by definition: $f_\text{L}(t) = \frac{1}{2\pi} \frac{\mathrm{d}\phi(t)}{\mathrm{d} t}$.

The beat signal arises from the interference between the local oscillator field (at time $t$) and the signal field reflected from a target with a round-trip delay $\tau$.
The resulting beat signal is governed by the phase difference between these two fields: $\Delta \phi(t) = \phi(t) - \phi(t-\tau)$.
Under the approximation that the delay is much shorter than the chirp duration ($\tau \ll T_{\text{chirp}}$), we can expand $\phi(t-\tau)$ using a first-order Taylor series around $t$:
$$\phi(t-\tau) \approx \phi(t) - \tau \frac{\mathrm{d}\phi(t)}{\mathrm{d}t}.$$
Substituting the frequency-phase relationship $\frac{\mathrm{d}\phi}{\mathrm{d}t} = 2\pi f_\text{L}(t)$ into this expansion yields the phase difference:
$$\Delta \phi(t) \approx \tau \frac{\mathrm{d}\phi(t)}{\mathrm{d}t} = 2\pi \tau f_\text{L}(t).$$
Consequently, the detected beat signal voltage is proportional to the cosine of this phase difference:
$$V(t) \propto \cos \left( \Delta \phi(t) \right) \approx \cos \left( 2\pi \tau f_\text{L}(t) \right).$$

With standard uniform time-domain sampling ($\Delta t = \text{const.}$), the nonlinear term $\delta \nu(t)$ introduces a non-uniform evolution of the cosine argument, leading to spectral broadening of the instrumental line shape, as shown in Supplementary Fig. \ref{SIfig:linearization}. 
To correct for this, we recast the signal domain from time $t$ to optical frequency $f_\text{L}$. 
In this domain, the signal implies:
$$V(f_\text{L}) \propto \cos \left( 2\pi \cdot \tau \cdot f_\text{L} \right).$$
This transformation renders the signal a pure sinusoid with a spectral oscillation frequency strictly determined by the delay $\tau$, independent of the chirp dynamics. 
Experimentally, we calibrate the instantaneous frequency using fiber ring cavities referenced to atomic hyperfine transitions. 
By resampling the interferograms onto a uniform optical frequency grid ($\Delta f_\text{L} = \text{const.}$), the nonlinearity $\delta \nu(t)$ is effectively compensated, restoring the transform-limited resolution of the system, as shown in Supplementary Fig. \ref{SIfig:linearization}.

\begin{figure}[h!]
\centering
\renewcommand{\figurename}{Supplementary Figure}
\includegraphics{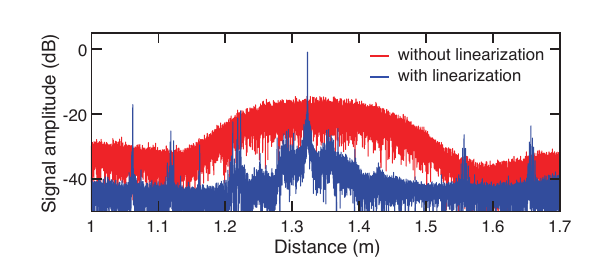}
\caption{
\textbf{Comparization of LiDAR range profile with and without linearization}. 
}
\label{SIfig:linearization}
\end{figure}
\clearpage

\section{Retrieval of gas transmission spectrum}
\vspace{0.5cm}
Following the frequency linearization described in Supplementary Note 6, the interferometric data is re-sampled onto a uniform optical frequency grid, $f_\text{L}$. 
This Note outlines the signal model and the processing workflow used to retrieve the normalized gas transmission spectrum, $T(f_\text{L})$.

The beat signal recorded by the photodetector arises from the heterodyne interference between the local oscillator (LO) field, $E_\text{LO}$, and the probe field, $E_\text{probe}$.
The probe field, having traversed the gas sample, carries the molecular absorption signature governed by the Beer-Lambert law: $|E_\text{probe}| = |E_\text{in}|\sqrt{T(f_\text{L})}$, where $E_\text{in}$ denotes the field incident on the gas cell.
Since both the LO and incident fields originate from the same laser source with a frequency-dependent output power $P_0(f_\text{L})$, their field amplitudes scale as $|E_\text{LO}| \propto |E_\text{in}| \propto \sqrt{P_0(f_\text{L})}$.
Consequently, the resulting beat signal---representing the interference cross-term---scales as:
$$V_\text{sig}(f_\text{L}) \propto \text{Re} \left[ E_\text{LO}^*(f_\text{L}) E_\text{probe}(f_\text{L}) \right] \propto P_0(f_\text{L})\sqrt{T(f_\text{L})}.$$
Simultaneously, a reference photodetector monitors the laser output intensity, yielding a baseline signal $V_\text{ref}(f_\text{L}) \propto P_0(f_\text{L})$.

The data processing workflow is illustrated in Supplementary Fig. \ref{SIfig:hilbert}.
To isolate the interference term from the low-frequency background and high-frequency noise, $V_\text{sig}(f_\text{L})$ is bandpass-filtered around the beat frequency. 
We then extract the instantaneous amplitude (envelope), $A(f_\text{L})$, via the Hilbert transform \cite{Marple:99}, $\mathcal{H}$:
$$A(f_\text{L}) = |V_\text{sig}(f_\text{L}) + i\mathcal{H}[V_\text{sig}(f_\text{L})]| \propto P_0(f_\text{L}) \sqrt{T(f_\text{L})}.$$
Finally, to retrieve the intensity transmission spectrum, the squared envelope is normalized by the squared reference signal. This operation effectively decouples the common-mode laser power fluctuations:
$$T_\text{meas}(f_\text{L}) = \frac{A^2(f_\text{L})}{[V_\text{ref}(f_\text{L})]^2} \propto \frac{P_0^2(f_\text{L}) \cdot T(f_\text{L})}{P_0^2(f_\text{L})} = T(f_\text{L}).$$
The derived spectrum is subsequently smoothed using a Savitzky--Golay filter \cite{Savitzky:64} to enhance the signal-to-noise ratio for spectroscopic fitting.

\begin{figure}[h!]
\centering
\renewcommand{\figurename}{Supplementary Figure}
\includegraphics{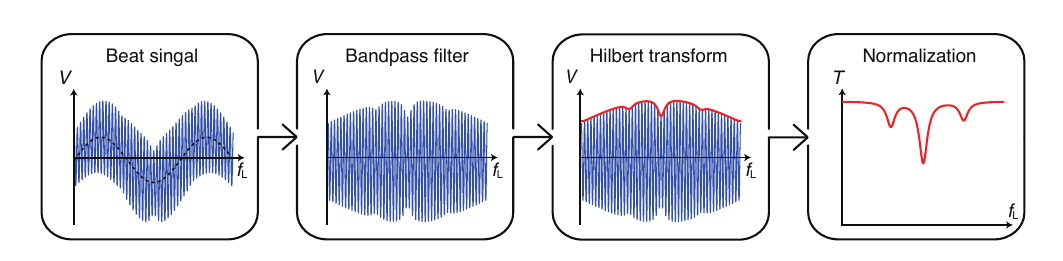}
\caption{
\textbf{Signal processing workflow}. 
}
\label{SIfig:hilbert}
\end{figure}

\section{Sum frequency generation in the visible band}
\begin{figure}[h!]
\centering
\renewcommand{\figurename}{Supplementary Figure}
\includegraphics{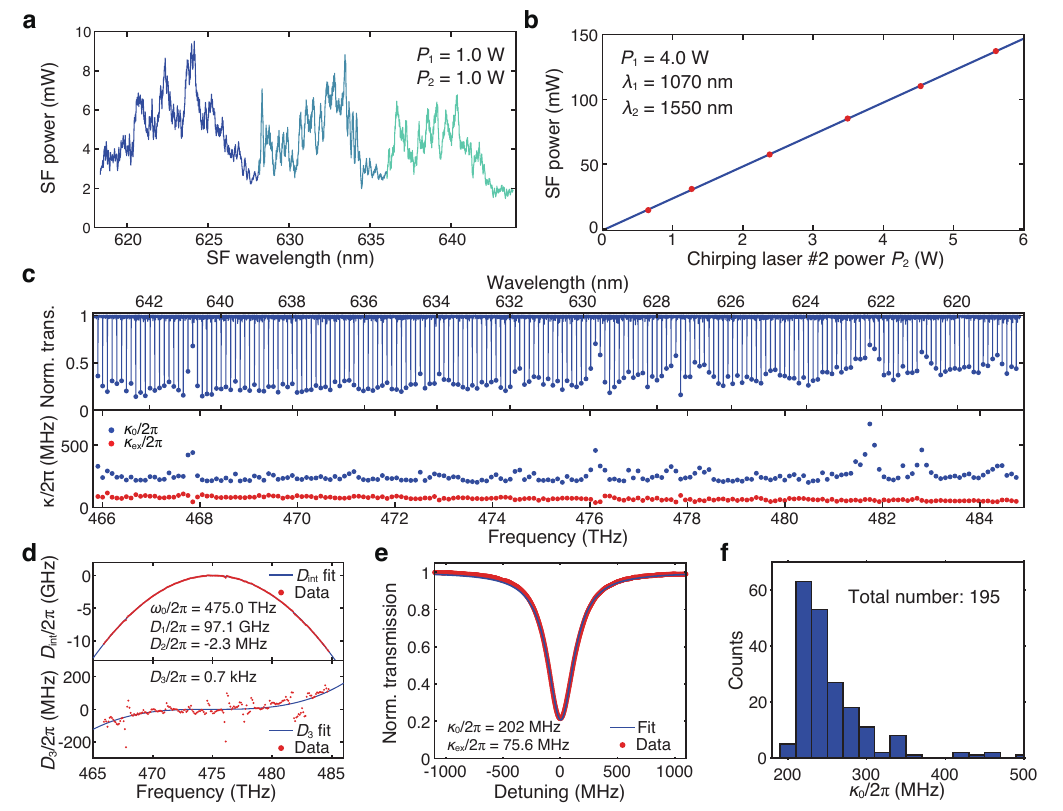}
\caption{
\textbf{Broadband SHG performance in the visible band and absolute frequency references}. 
\textbf{a}.
Full spectral coverage characterization. 
By stepping the start wavelength of chirping laser \#1 (1035, 1057, and 1079 nm), the system achieves seamless coverage from 618 to 644 nm. 
These tuning curves are characterized with input powers set at chirping laser \#1 power $P_1=1.0$ W and chirping laser \#2 power $P_2=1.0$ W.
\textbf{b}.
Power transfer measurement. 
The SF output power is plotted against $P_2$ (fixed parameters: $\lambda_1=1070$ nm, $\lambda_2=1550$ nm, $P_1=4.0$ W). 
A maximum SF power of 137 mW is generated at $P_2=5.6$ W.
\textbf{c}.
Transmission spectra of a silicon nitride microresonator from 618 to 644 nm.
\textbf{d}.
Measured microresonator's integrated dispersion profile and fit up to the third order $D_3$.
The deviation is defined as $[D_\text{int}(\mu)-D_2\mu^2/2]/2\pi$, showing a clear $D_3$ trend.
\textbf{e}.
Measured transmission of one under-coupled resonance at 472.2 THz.
\textbf{f}.
Histogram of measured $\kappa_0/2\pi$ values of the 195 resonances from one microresonator, showing the most probable value of $\kappa_0/2\pi=220$ MHz.
}
\label{SIfig:SFG}
\end{figure}

\vspace{0.5cm}
In addition to DFG, sum frequency generation (SFG) can also be utilized via CPLN waveguide. 
The CPLN waveguide features a top width of 7~\textmu m, an etching depth of 5~\textmu m, a total thickness of 10~\textmu m, and a sidewall angle of 70$^{\circ}$. 
The total device length is 20~mm.
To satisfy the CQPM condition over a broad spectral range, the poling period ($\Lambda$) is linearly chirped from 10.97~\textmu m to 12.47~\textmu m along the propagation direction and the duty cycle ($\eta$) follows a tanh-apodized profile with $\beta$ = 6.
By synchronously chirping laser \#1 and chirping laser \#2 in the same spectral direction, we synthesized sum frequency (SF) scans spanning 618 to 644 nm with 7.49 THz mode-hop-free tuning bandwidth (Supplementary Fig. \ref{SIfig:SFG}a).
At pump parameters of $\lambda_1=1070$ nm ($P_1=4.0$ W) and $\lambda_2=1550$ nm ($P_2=5.6$ W), the system delivers a maximum output power of 137 mW (Supplementary Fig. \ref{SIfig:SFG}b).

The SF laser is subsequently employed to characterize a Si$_3$N$_4$ microresonator. 
Its normalized transmission spectrum (618--641 nm) is presented in Supplementary Fig. \ref{SIfig:SFG}c (top). 
A typical resonance fit, shown in Supplementary Fig. \ref{SIfig:SFG}e, revealing $\kappa_0/2\pi = 202.0$~MHz and $\kappa_\mathrm{ex}/2\pi = 75.6$~MHz, respectively. 
We performed Lorentzian fitting for all resonances within the scan range; 
the extracted $\kappa_0/2\pi$ and $\kappa_\mathrm{ex}/2\pi$ values are plotted in Supplementary Fig. \ref{SIfig:SFG}c (bottom). 
Statistical analysis in Supplementary Fig. \ref{SIfig:SFG}f yields a most probable value of $\kappa_0/2\pi = 220$~MHz, corresponding to $Q_0 = 2.2 \times 10^6$ and a linear propagation loss of 43~dB/m. 
Additionally, the microresonator dispersion was characterized up to the third-order term ($D_3$), as detailed in Supplementary Fig. \ref{SIfig:SFG}d.

\clearpage

\vspace{1cm}
\section*{Supplementary References}
\bibliographystyle{apsrev4-1}
\bibliography{bibliography}